\documentclass[a4paper,fleqn,usenatbib]{mnras}

\usepackage{newtxtext,newtxmath}

\usepackage[T1]{fontenc}
\usepackage{ae,aecompl}

\usepackage{graphicx}	
\usepackage{amsmath}	

\title[Class I YSO outbursts]{On the incidence of episodic accretion in Class I YSOs from VVV}

\author[C. Contreras Pe\~{n}a et al.]{
Carlos Contreras Pe\~{n}a,$^{1,2,3}$\thanks{E-mail: ccontreras@snu.ac.kr (CCP)}
Philip W. Lucas,$^{1}$,
Zhen Guo$^{4,5,6,1}$, 
and Leigh Smith$^{7}$
\\
$^{1}$Centre for Astrophysics Research, University of Hertfordshire, College Lane, Hatfield, AL10 9AB, UK\\
$^{2}$Department of Physics and Astronomy, Seoul National University, 1 Gwanak-ro, Gwanak-gu, Seoul 08826, Republic of Korea\\
$^{3}$Research Institute of Basic Sciences, Seoul National University, Seoul 08826, Republic of Korea\\
$^{4}$Instituto de F{\'i}sica y Astronom{\'i}a, Universidad de Valpara{\'i}so, ave. Gran Breta{\~n}a, 1111, Casilla 5030, Valpara{\'i}so, Chile\\
$^{5}$N\'ucleo Milenio de Formaci\'on Planetaria (NPF), ave. Gran Breta{\~n}a, 1111, Casilla 5030, Valpara{\'i}so, Chile\\
$^{6}$Departamento de F{\'i}sica, Universidad Tecnic{\'a} Federico Santa Mar{\'i}a, Avenida Espa{\~n}a 1680, Valpara{\'i}so, Chile
\\
$^{7}$Institute of Astronomy, University of Cambridge, Madingley Road, Cambridge CB3 0HA, UK}

\date{Accepted XXX. Received YYY; in original form ZZZ}

\pubyear{2023}

\begin{document}
\label{firstpage}
\pagerange{\pageref{firstpage}--\pageref{lastpage}}
\maketitle

\begin{abstract}
Episodic accretion is one of the competing models to explain the observed luminosity spread in young stellar clusters. These short-lived high accretion events could also have a strong impact on planet formation. Observations of high-amplitude variability in young stellar objects (YSOs) due to large changes in the accretion rate provide direct observational evidence for episodic accretion. However there are still uncertainties in the frequency of these events and if episodic accretion is universal among YSOs. To determine the frequency of outbursts in Class I YSOs, we built a large and robust sample of objects at this evolutionary stage, and searched for high-amplitude near-infrared ($\Delta K_{\rm S}>2$~mag) variability in the VIRAC2 database of the Vista Variables in the Via Lactea (VVV) survey. By complementing with near-IR (2MASS and DENIS) and mid-IR (WISE/Neo-WISE) data, we find that from $\sim$ 7000 Class I YSOs, 97 objects can be classified as eruptive variable YSOs. The duration of the outbursts vary from a few months to longer than 9 years, and cover a similar range of amplitudes. Values of $\Delta K_{\rm S}>5$~mag, however, are only observed in outbursts with duration longer than 9 years. When considering different effects of completeness and contamination we estimate that the incidence of episodic accretion in Class I YSOs is between 2\% and 3\%. Finally, we determine a recurrence timescale of long-term outbursts (a.k.a FUors) of $\tau=1.75^{+1.12}_{-0.87}$~kyr. The latter value agrees with previous estimates and is in line with the expectations of higher frequency of FUor outbursts during younger stages of evolution.
\end{abstract}

\begin{keywords}
stars: formation -- stars: protostars -- stars: pre-main-sequence -- stars: variables: T Tauri, Herbig Ae/Be 
\end{keywords}



\section{Introduction}\label{sec:intro}

In the episodic accretion model of star formation, stars gain most of their mass during short-lived episodes of high accretion ($\dot{\mathrm{M}}\simeq10^{-4}$ M$_{\odot}$ yr$^{-1}$) followed by long periods of quiescent low-level accretion \citep{1996Hartmann}. Here, instabilities in the disc lead to sudden episodes of enhanced accretion. There are many models competing to explain the instabilities that give rise to the outbursts, and it is not yet clear which physical mechanism governs the transport of angular momentum. The models include gravitational and magnetorotational instabilities \citep{2009Zhu, 2020Kadam}, thermal viscous instability \citep{1994Bell,1995Bell}, disc fragmentation \citep{2015Vorobyov}, binary interaction \citep{2010Reipurth}, stellar flybys \citep{2019Cuello,2022Dong} and planet-disc interaction \citep{2004Lodato}. 

Young stellar objects (YSOs) display sudden rises in brightness that provide direct observational evidence for episodic accretion \citep{1996Hartmann}. YSO outbursts are usually divided according to the outburst duration ($t_{out}$) and spectroscopic characteristics during outburst \citep{2018Connelley}, with the original classification scheme arising from observations at optical wavelengths. Traditionally, the outbursts were classified into FUors ($t_{out}>10$ yr) and EX Lupi-type ( $t_{out}<1$~yr). EX Lupi-type outbursts show strong emission from hydrogen recombination lines and from first overtone CO bands (at 2.3--2.4 $\mu$m). On the other hand, FUors display a lack of emission lines and strong first overtone CO absorption. The lack of emission lines is unexpected given the strong correlation between H I emission and mass accretion rate observed in YSOs due to magnetospheric accretion \citep{1998Muzerolle_a,2004Muzerolle}. This could be explained if at the high accretion rates of FUor outbursts ($\dot{\mathrm{M}}=10^{-4}$ M$_{\odot}$ yr$^{-1}$) the ram pressure from the disc prevents the formation of funnel flows along field lines and the mechanism from which H I lines arise \citep[``crushing of the magnetosphere''][]{2012Fischer}. In the classical interpretation of eruptive YSOs, FUor outbursts were thought to be characteristic of Class I of flat-spectrum YSOs whilst EX Lupi-type outbursts were associated with the older Class II YSOs \citep[e.g.][]{2001Sandell}.

More recent discoveries of YSO outbursts tend to blur the original classification scheme and interpretation \citep{2014Audard, 2022Fischer}. This classification now appears obsolete
with regard to the duration of EX Lupi-type outbursts because spectroscopy of the
large VVV sample of eruptive YSOs \citep{2017Contreras, 2021Guo} has shown
that EX Lupi-type events predominate even amongst long duration outbursts
\citep{2021Guo}. In addition, some YSOs with FUor spectra have smaller amplitude and short duration, such as V322 in \citet{2017Contreras} and \citet{2020Guo}. Finally, FUor outbursts do occur during the Class II stage \citep{2014Gramajo, 2019Contreras}.

Episodic accretion has been invoked to solve some long-standing problems in star formation. The discrepancy between the observed and predicted luminosities of  Class I YSOs  \citep{1990Kenyon,2009Evans} supports the idea that stars spend most of their time at low accretion states. The outbursts can also have a long-lasting impact on the properties of the central star such as luminosity and radius \citep{2017Baraffe} and could explain the observed spread in Hertzprung-Russell diagrams of pre-main-sequence clusters. The fact that accretion is episodic can have implications for many aspects of star and planet formation. The long periods spent at low accretion stages would allow the disc to cool sufficiently to fragment, helping to produce low-mass companions \citep{2012Stamatellos}. Outbursts have an impact on processes of planet formation as they alter the chemistry of protoplanetary discs \citep{2019Artur}, the location of the snowline of various ices \citep{2016Cieza} and could affect orbital evolution of planets  \citep{2013Boss,2021Becker}. The observations of calcium--aluminium-rich inclusions in chondrites \citep{2009Wurm},  and the depletion of lithophile elements \citep{2014Hubbard} and refractory carbon in Earth \citep{2018Klarmann} could be evidence for past large eruptions in our own Solar system.

Determining the incidence of high-accretion events in the observed population at various stages of early stellar evolution and the implied interval between events
(if the phenomenon is universal) is a crucial step to inform future theoretical models that aim to study the impact of episodic accretion on stellar and/or planetary formation.
 \citet{2015Hillenbrand} state that to determine the frequency of events from observations, it is necessary to maximise both the number of YSOs surveyed and the time baseline of such observations. Previous efforts to determine the frequency of outbursts show that these events are more common during the earlier stages of evolution \citep{2012Ioannidis, 2013Scholz, 2019Fischer, 2019Hsieh, 2019Contreras,2021Park, 2022Zakri}. However, the estimated intervals during the Class I stage arise from observations of either small samples of YSOs or short timescales, which yield large uncertainties.

In this paper, we present the results of the study we conducted using the multi-epoch photometry from the Vista Variables in the Via Lactea Survey  and its extension (VVV/VVVX) for a large sample of YSOs. It is our aim to answer the questions of outburst amplitudes and frequencies during the Class I stage on evolution. The paper is divided as follows: Section \ref{sec:sample} describes the sample of YSOs used in this work, as well the method to classify them into the Class I stage. In addition we describe the photometric data from the VVV survey. In Section \ref{sec:method} we describe the search of high-amplitude variable stars in the VIRAC catalogue, as well as the initial classification from visual inspection of light curves. In Section \ref{sec:acc_out} we determine the sample of objects where variability is more likely to be driven by changes in the accretion rate. In here we also divide them according to the duration of the outbursts. From this sample we study the incidence of episodic accretion during the Class I stage in Section \ref{sec:inc_out}. In Section \ref{sec:taucalc} we determine the frequency of long-term outbursts (a.k.a FUors) during the Class I stage. Finally we present a summary of our findings in Section \ref{sec:sum}.

\section{Data}\label{sec:sample}

The long baseline near-infrared observations from the VVV/VVVX survey as well as publicly available YSO catalogues are used to maximise the requirements to measure the incidence of episodic accretion, i.e. the sample of Class I YSOs and the monitoring baseline \citep{2015Hillenbrand, 2019Contreras}. The data used in this work is summarised below.

\subsection{The VVV/VVVX survey}

The data for the VVV/VVVX surveys were collected by the Visible and Infrared Survey Telescope for Astronomy (VISTA) 4m telescope located at Cerro Paranal Observatory in Chile. VISTA is equipped with VIRCAM, a near-infrared camera with a 4$\times$4 array of 2048$\times$2048 Raytheon Virgo HgCdTe pixel detectors, and a typical pixel scale of 0.\arcsec 339.  The array has large spacing along the X and Y axis with each detector covering 694$\times$694 arcsec$^{2}$. A single pointing (or ``pawprint'') provides partial covering over 0.59 deg$^{2}$, whilst  continuous coverage of a particular field is achieved by combining six single pointings with appropriate offsets. This combined image is called a tile. The images are combined and processed at the Cambridge Astronomical Survey Unit (CASU).

 The regions covered by the VVV survey comprised the Galactic Bulge region within $-10^{\circ} < l < +10^{\circ} $ and $-10^{\circ} < b < +5^{\circ}$ and the Galactic Disc region within $295^{\circ} < l < 350^{\circ} $ and $-2^{\circ} < b < +2^{\circ}$. These regions were observed using filters $Z(\lambda_{\rm eff}=0.88~\mu$m), $Y(\lambda_{\rm eff}=1.02~\mu$m), $J(\lambda_{\rm eff}=1.25~\mu$m), $H(\lambda_{\rm eff}=1.65~\mu$m) and $K_{\rm s}(\lambda_{\rm eff}=2.15~\mu$m). Each region was observed with two epochs of contemporaneous JHK$_{\rm s}$  in 2010 and 2015, and separately two contemporaneous epochs of ZY in 2010 and 2015.

 The survey was extended (as VVVX) to cover additional areas in the Galactic Bulge and Disc as well as to provide 9 additional epochs of K$_{\rm s}$ photometry in the areas already covered by the original strategy from VVV.  The observations in the original VVV regions were carried out between
2016 and 2019. Observations of the new areas were completed in early 2022, but these
areas are not analysed in this work.

\subsection{The sample}\label{ssec:sample}

We created a list of YSOs from objects found in the \citet{2008Robitaille}, \citet{2016Marton} and \cite{2020Kuhn} catalogues of young stars. A brief description of these catalogues is presented below.

\begin{description}

\item {\bf \citet{2020Kuhn}}  The {\it Spitzer}/IRAC Candidate YSO Catalog for the Inner Galactic Midplane (SPICY) uses a random forest classification to select YSOs using {\it Spitzer} measurements of the 3 to 24 $\mu$m spectral energy distribution obtained during the cryogenic mission. This includes seven {\it Spitzer}/IRAC surveys covering 613 square degrees. The data is also augmented with near-infrared surveys 2MASS \citep{2006Skrutskie}, UKIDSS Galactic Plane Survey \citep{2007Lawrence,c2008Lucas} and VVV \citep{2012Saito}. The SPICY catalogue contains 62959 candidate YSOs that fall in the region covered by the VVV survey. 

\item {\bf \citet{2008Robitaille}}  The authors searched for objects with intrinsically red mid-IR colours based on observations from the {\it Spitzer}/IRAC Glimpse I and II surveys \citep{2009Churchwell}. The analysis aimed to provide a reliable source catalogue. Given this, the work is based on photometry obtained from the Glimpse Point Source Catalogues, which are more reliable than Point Source archives. In addition, \citet{2008Robitaille} selected sources with high Signal-to-Noise detections that fulfilled 13.89$\geq$[4.5] and $9.52\geq$[8.0]. \citet{2008Robitaille} performs visual inspection of red sources and PSF photometry to compare with the photometry from Glimpse catalogues. When found, uncertain photometric values from the catalogues are replaced by PSF measurements.

The red source catalogue is built by selecting sources with $[4.5]-[8.0]\geq1$. This population is comprised of a variety of objects, including Planetary Nebulae (PNe), Extragalactic sources, Asymptotic Giant Branch (AGB) stars and YSOs.  The mid-IR colours are used by \citet{2008Robitaille} to discriminate between the different populations, where Extragalactic sources and PNe only account for small percentage of the red catalogue ($<3$\%). 

The remaining 97\% sources in the red catalogue are classified as either candidate YSOs or candidate AGB stars, based on their magnitude and colours. Carbon- and oxygen-rich AGB stars with high mass loss \citep[``xAGBs'' in][]{2008Robitaille} are likely the to represent the majority of AGB stars in the red source catalogue, as they tend to be much redder than standard AGB stars (or ``sAGBs''). Most sAGBs are bluer than the limit of $[4.5]-[8.0]=1$ used to select red sources in \citet{2008Robitaille}. Since AGB stars are not distinguishable from the whole population of red sources in the IRAC colour-colour space, \citet{2008Robitaille} includes MIPS 24 $\mu$m data as AGB stars tend to have bluer $[8.0]-[24]$ colours on average than the overall red population. xAGBs also account for the brighter and bluer peak seen in $[4.5]$ vs $[4.5]-[8.0]$ colour-magnitude diagrams. Given this, \citet{2008Robitaille} classifies objects with $[4.5]$ brighter than 7.8 mag as xAGB candidates. Objects fainter than this limit, but which show $[8.0]-[24]$ colour bluer than 2.5 mag are classified as sAGB candidates. Objects in the red catalogue that are not extragalactic sources, PNe or AGB stars are therefore classified as candidate YSOs. There are 12103 candidate YSOs from the red catalogue in the area covered by the VVV survey.

\item {\bf \citet{2016Marton}}. Through the use of a supervised learning algorithm \citet{2016Marton} classify objects as YSOs based on 2MASS $JHK_{\rm s}$ and {\it WISE} \citep{2010Wright} photometry. The authors provide two tables with YSOs that are either classified as Class III or Class I/II YSOs. Since we are interested in the younger sources we use the latter table where we find 57516 candidate YSOs in the area covered by the VVV surveys.

\end{description}

The combination of the three catalogues (without accounting for possible repetition) yield 132578 candidate YSOs in the VVV area.  

\subsection{YSO Class}\label{ssec:class}

The observed SED of young stellar objects is generally used to determine the likely evolutionary stage of the system \citep[e.g.][]{1987Lada, 1994Greene, 2014Dunham}. The different YSO Classes (which are thought to be associated with different evolutionary stages) can be defined according to the value of their infrared spectral index, $\alpha$, defined as

\begin{equation}
\alpha = \frac{dlog(\lambda F_{\lambda})}{dlog\lambda},
\label{eq:sed}
\end{equation}

\noindent where in the original classification from \citet{1987Lada}, estimated from the observed SED between 2 $< \lambda <$ 20 $\mu$m, objects with positive spectral indices ($\alpha>0$) are still embedded in an infalling envelope and are classified as Class I YSOs. Objects where the envelope had dissipated show negative spectral indices ($-2<\alpha<0$) and are defined as Class II YSOs. An additional classification was added to include objects which emit most of their radiation at $\lambda>10 ~\mu$m \cite[Class 0 YSOs][]{1993Andre}, which are believed to be at a younger evolutionary stage than Class I YSOs. Finally, \citet{1994Greene} defines YSOs with $-0.3<\alpha<0.3$ as ``Flat-Spectrum'' YSOs, that are likely transition objects between the Class I and Class II stages. 

Since the value of $\alpha$ estimated using eqn \ref{eq:sed} can be affected by reddening \citep[see e.g.][]{2020Kuhn}, 4.5$-$24 or $3.4-22 ~\mu$m colours can be used to derive $\alpha$ \citep[see e.g.][]{2017Kang,2020Kuhn}.

YSOs are also classified into the different stages of evolution based on mid-IR colour-colour diagrams \citep[see e.g.][]{2009Gutermuth,2014Koenig}. Based on colours of known YSOs, the 3.4--12 $\mu$m wavelength range is used to define a criterion to classify objects as either Class I or Class II YSOs. In addition, the criterion is used to discard possible contamination from evolved stars, unresolved shock emission knots and objects that suffer from structured PAH aperture contamination. In the case of the Class I criteria of \citet{2009Gutermuth}, additional photometry at 24 $\mu$m is used to define possible contamination from reddened Class II YSOs.

Interestingly, during the analysis of this work we noted that sources that are classified as Class I YSOs from either the \citet{2009Gutermuth} or \citet{2014Koenig} criteria tend to include objects with negative values of $\alpha$ (determined from mid-IR colours), but generally larger than $\alpha=-0.3$. YSOs with these values of $\alpha$ could be considered as transition or flat-spectrum sources using the definition of \citet{1994Greene}.

It is the main objective of this study to determine the outburst incidence/rate during the embedded stage of YSO evolution (also known as Class I YSOs). In the following, we describe the criteria used to determine a reliable sample of YSOs that show colours and SEDs consistent with those of Class  I YSOs. 

\begin{description}

\item {\bf (i)} From the sample of 132578 candidate YSOs, we initially select only objects with detection at 22 or 24 $\mu$m in {\it WISE} or {\it Spitzer}. This is done to ensure that YSOs have rising SEDs towards longer wavelengths.

\item {\bf (ii)} From objects that fulfil (i), we used the short-wavelength (3.4--12 $\mu$m) photometry in {\it WISE} or {\it Spitzer}, and the classification criteria of  \citet{2009Gutermuth} or \citet{2014Koenig}, to establish likely Class I YSOs whilst discarding possible contamination from unresolved shock emission knots and structured PAH aperture contamination.

\item {\bf (iii)} Finally, for all of the objects selected in (i) and (ii), the $3.4-22 ~\mu$m or $4.5-24 ~\mu$m colours are used to determine the value of the spectral index, $\alpha_{24}$. The final catalogue to be used in the search for eruptive variable YSOs contains all YSOs with $\alpha_{24}>-0.3$. 

\end{description}  
We apply these steps to classify the YSOs in the three catalogues. We find that 5541, 2580 and 1190 objects fulfil the criteria in  the SPICY, \citet{2008Robitaille} and \citet{2016Marton} catalogues, respectively.

We crossmatched the Class I YSOs obtained from each catalogue to avoid repetition of sources. Not surprisingly, most of the red objects from \citet{2008Robitaille}, 1917 YSOs, are contained within SPICY. Therefore there are additional 663 Class I YSOs from \citet{2008Robitaille}. These objects are either not classified as YSOs in SPICY or the {\it Spitzer} catalogues lack photometry at some of the short-wavelength filters. Since \citet{2008Robitaille} perform their own photometry, these objects have available photometry at the filters used to classify YSOs with the \citet{2009Gutermuth} criteria. Finally, 189 sources arising from the \citet{2016Marton} catalogue are found in SPICY or \citet{2008Robitaille}, these leaves 1001 additional Class I YSOs that are only found in the \citet{2016Marton} catalogue.

In summary, we were able to compile a list of 7205 Class I YSO candidates. Table \ref{tab:ysos} shows the summary data for all the YSOs selected from the different samples described above. In the table we include the designation of the YSO, right ascension, declination, $\alpha_{24}$ and the catalogue where the YSO comes from.

\begin{table}
	\centering
	\caption{Summary of Class I YSOs. The full version of this table will be made available online.}
	\label{tab:ysos}
\resizebox{\columnwidth}{!}{
   	\begin{tabular}{lcccc} 
		\hline
		ID & RA & DEC & $\alpha_{24}$ & source\\
		\hline
            YSO1    & 11:42:08.1 & $-$62:37:45.0 &  -0.1 & SPICY \\
            YSO2    & 11:42:43.5 & $-$62:21:14.6 &   0.1 & SPICY \\
            YSO3    & 11:42:56.2 & $-$62:37:24.8 &   1.1 & SPICY \\
            YSO4    & 11:43:09.3 & $-$62:27:19.9 &   0.2 & SPICY \\
            YSO5    & 11:43:19.9 & $-$62:25:34.9 &   0.9 & SPICY \\
            YSO6    & 11:43:24.2 & $-$62:29:51.7 &  -0.2 & SPICY \\
            YSO7    & 11:43:44.7 & $-$62:36:47.1 &   0.6 & SPICY \\
            YSO8    & 11:44:12.9 & $-$62:34:49.2 &  -0.1 & SPICY \\
            YSO9    & 11:44:19.1 & $-$62:38:21.2 &   0.7 & SPICY \\
            YSO10   & 11:44:36.3 & $-$62:40:05.6 &   0.6 & SPICY \\
            YSO11   & 11:44:38.3 & $-$62:13:14.1 &   0.5 & SPICY \\
		\hline
	\end{tabular}
}
\end{table}

\section{VIRAC search}\label{sec:method}

In this work, we use a preliminary version of the VVV/VIRAC2 catalogue (\citealt{2018Smith}, Smith et al. in prep). The catalogue contains the Z, Y, J  H and K$_{\rm s}$ profile fitting photometry from the 2010 to 2019 VVV/VVVX observations for $\sim$700 million sources, along with proper motions and parallaxes computed with the K$_{\rm s}$ data. The PSF photometry is derived using DoPHOT \citep{1993Schecter,2012Alonso} where a new absolute photometric calibration is obtained to mitigate issues that arise from blending of VVV sources in 2MASS data of crowded inner Galactic bulge regions (\citealt{2020Hajdu}, Smith et al. in prep).

The VIRAC2 catalogue also provides variability indices that are useful to distinguish real variable stars from false positives: the Stetson I index \citep{1993Welch, 1996Stetson} and the von Neumann Eta index \citep{1941Neumann}. It also contains several parameters that characterise the K$_{\rm s}$ magnitude distribution of the light curve by percentiles, e.g. K$_{{\rm s},p0}$, K$_{{\rm s}, p4}$, K$_{{\rm s}, p50}$, K$_{{\rm s}, p96}$ or K$_{{\rm s}, p100}$.

We crossmatched the catalogue of Class I YSOs with the VIRAC2 catalogue using a 1\arcsec~radius. We find 5661 Class I YSOs with VIRAC2 counterparts. To avoid selecting sources where high-amplitude variability is determined by only 1 or 2 points in the light curve, we defined $\Delta K_{{\rm s}, p96}$ as the 96th minus 4th percentile in magnitude. \citet{2017Contreras_a} find that YSOs where variability can be explained by physical processes other than episodic accretion tend to show $\Delta K_{\rm s}<2$~mag. To avoid a high-contamination from these type of variable YSOs, candidates for eruptive variability are obtained from selecting sources with amplitudes, $\Delta K_{{\rm s}, p96}$, larger than 2 mag. This amplitude is also consistent with the expected change in brightness at near-IR wavelengths due to accretion outbursts \citep[][]{2022Hillenbrand}. This threshold does not necessarily ensure that we are selecting a clean sample of eruptive YSOs, as extinction-driven changes can show these high amplitudes. In addition, we might lose accretion-driven events of lower amplitudes, but these are likely less significant with respect to the mass accretion history and the radiative feedback on the accretion disc. We investigate the effects of the selected amplitude cut in Section \ref{sec:inc_out}. 

The amplitude cut of $\Delta K_{{\rm s}, p96}>2$~mag yields 304 candidate eruptive YSOs.

\subsection{Classification}\label{ssec:hamp_class}

High-amplitude variability in the near-IR is not only observed in eruptive variables. Other physical mechanisms, such as obscuration events due to inhomogeneities in the disc, can drive large amplitude variability. In addition, AGB stars, which can contaminate YSO catalogues, also vary with large amplitudes \citep{2017Contreras, 2021Guo}. Inspection of long-term, multi-wavelength light curves can help to discriminate between the different variability types. 

Additional data was obtained from near- and mid-IR public catalogues available from Vizier \citep{2000Oschsenbein} and the NASA/IPAC Infrared Science Archive (IRSA). These catalogues include 2MASS \citep{2006Skrutskie}, DENIS \citep{1994Epchtein}, UKIDSS GPS \citep{c2008Lucas}, K$_{\rm s}$ observation of G305.2 \citep{2007Longmore}, Spitzer and WISE/NEOWISE surveys \citep{2003Benjamin, 2010Wright}. For each candidate, the single exposure source databases from {\it WISE} and  {\it NEOWISE} surveys were queried using a 3\arcsec radius. {\it WISE} fluxes are corrected for saturation following the guidance from the {\it WISE} supplementary material \citep{2012Cutri}\footnote{https://wise2.ipac.caltech.edu/docs/release/neowise/expsup/}. To provide a better comparison between mid-IR surveys, {\it Spitzer} photometry from the IRAC1 and IRAC2 filters was converted to {\it WISE} $W1$ and $W2$ using the equations from \citet{2014Antoniucci}.

Near- and mid-IR 3$\times$3 \arcmin cutout images from VVV, 2MASS, UKIDSS GPS, {\it Spitzer} and {\it WISE} were also inspected to confirm the high-amplitude variability in our candidates.

The light curve inspection reveals that the variability of our sample shows a wide range in behaviour. The 304 high-amplitude variables are initially classified into the light curve categories (or a combination of them) defined in \citet{2017Contreras}. These categories are defined in \citet{2017Contreras} as

\begin{description}

\item {\it Dippers} Objects found at approximately constant brightness that show sudden fading events lasting from months to years followed by a return to the previous brightness level. 

\item {\it Faders} Objects found at approximately constant brightness that start to fade over long timescales. 

\item {\it Short-term variables (STVs)} Objects that show apparent irregular variability with timescales less than $\sim$ 1 year. 

\item{\it Periodic} Objects showing periodic variability with periods from months to longer than 1 year.

\item{\it Eruptive--ccp17} Objects that show lightcurves clearly resembling those of the known subclasses of young eruptive variables. In general, in this first classification attempt, this class would include objects with outbursts lasting more than 1 year. We note that the ccp17 suffix is added to avoid confusion later on in the classification of accretion-driven outbursts.

\item{\it Unclassified} Objects that do not show a distinctive lightcurve that could be classified in the above classes.

\end{description}

\begin{figure*}
\includegraphics[width=\columnwidth,angle=0]{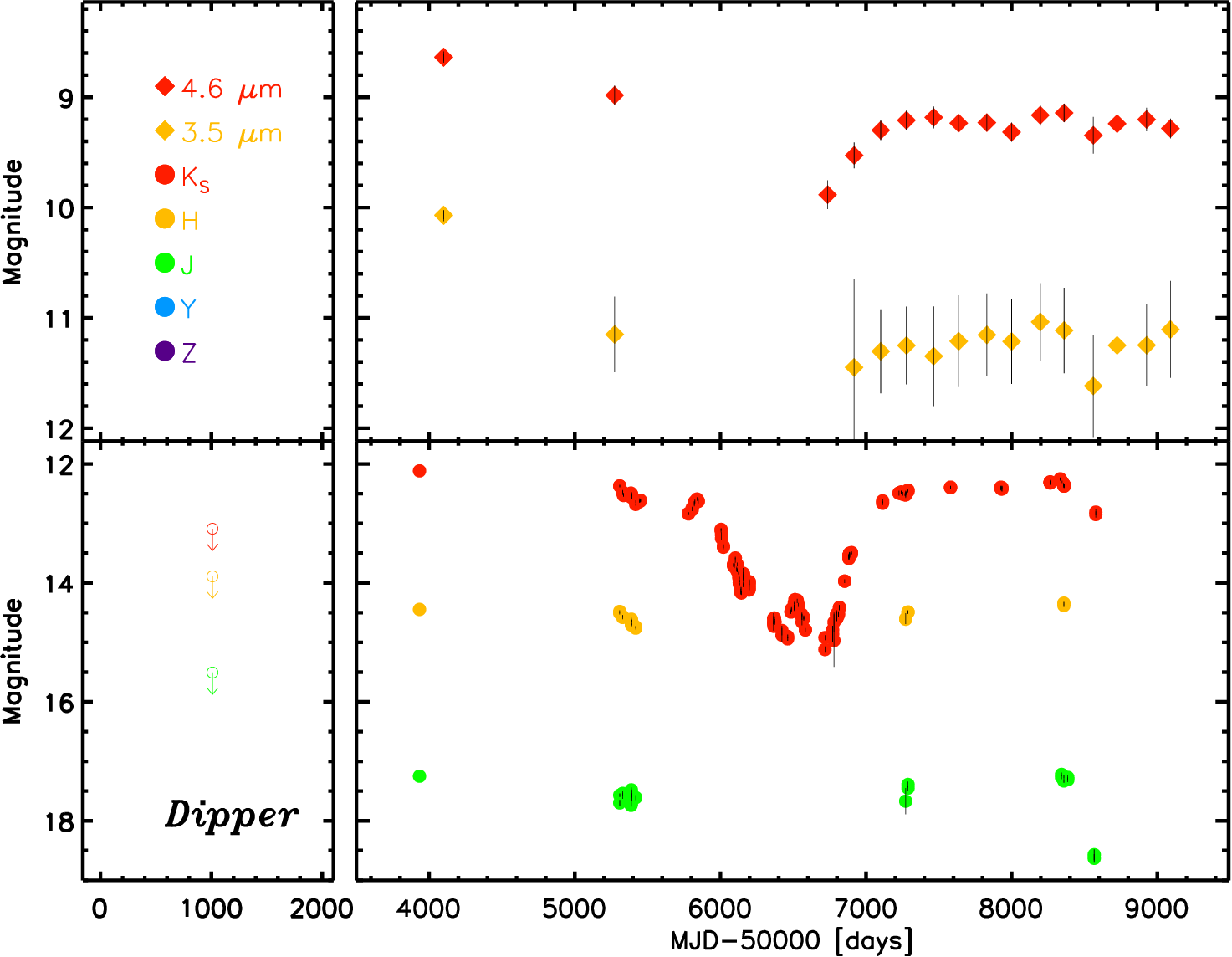}
\includegraphics[width=\columnwidth,angle=0]{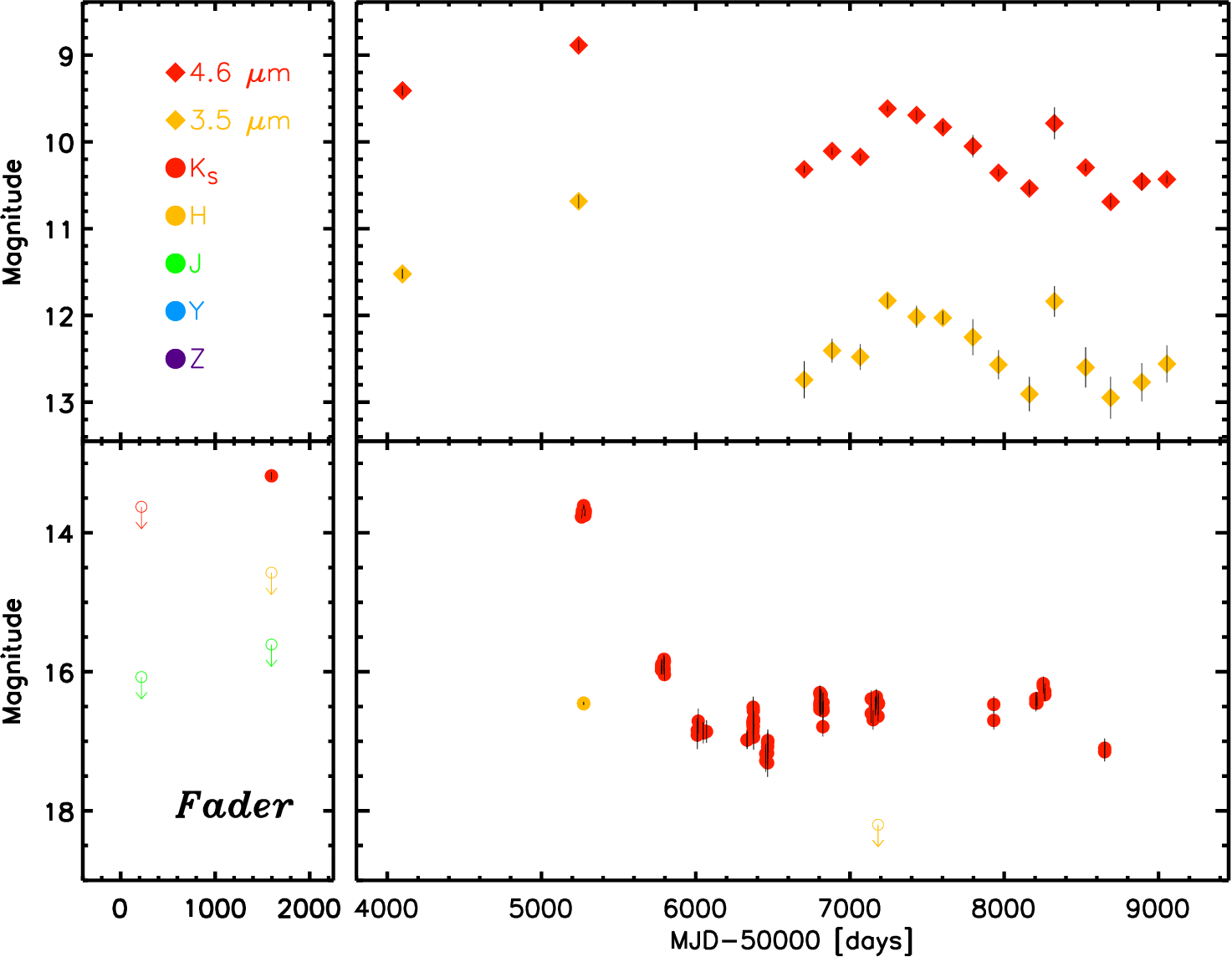}\\
\includegraphics[width=\columnwidth,angle=0]{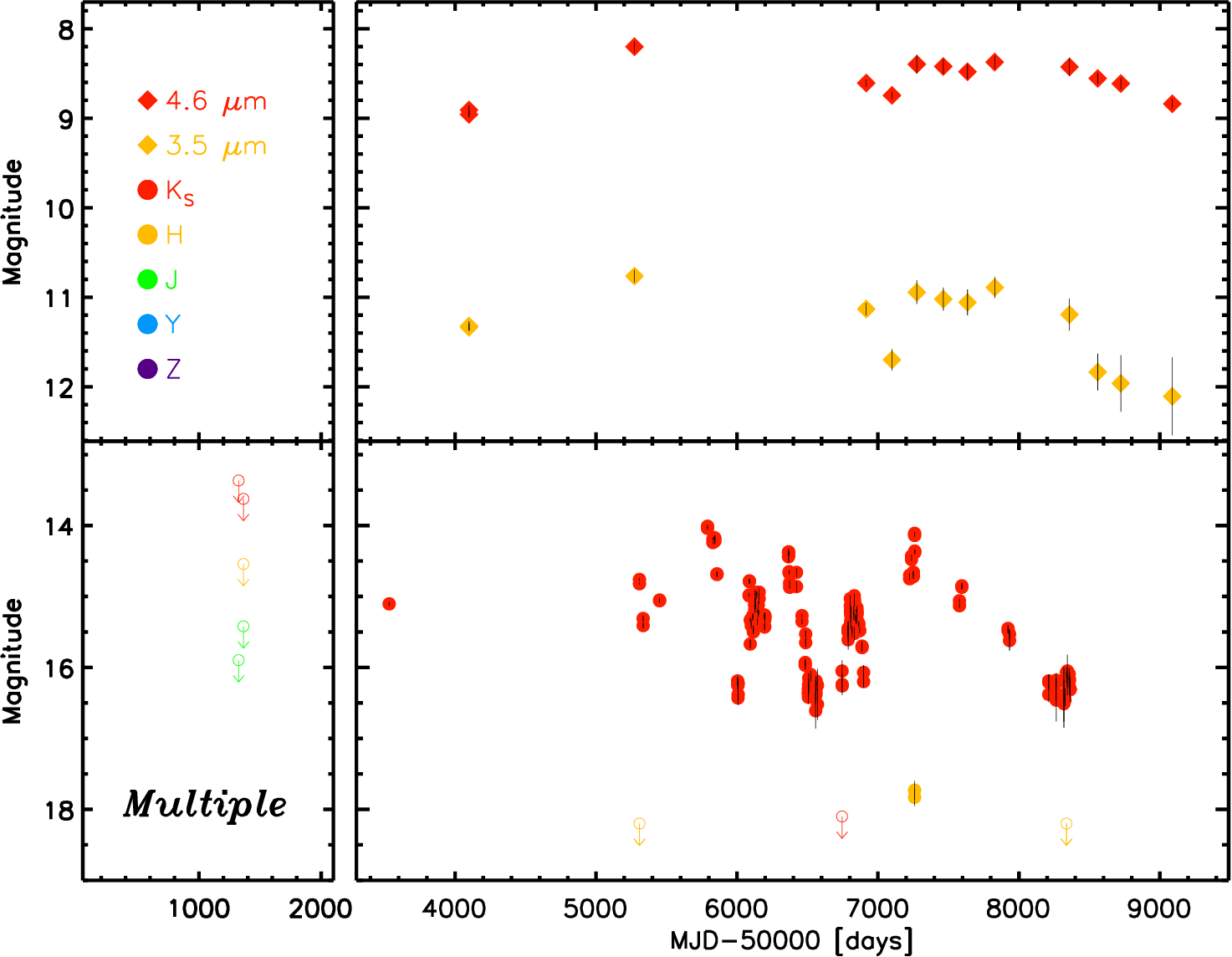}
\includegraphics[width=\columnwidth,angle=0]{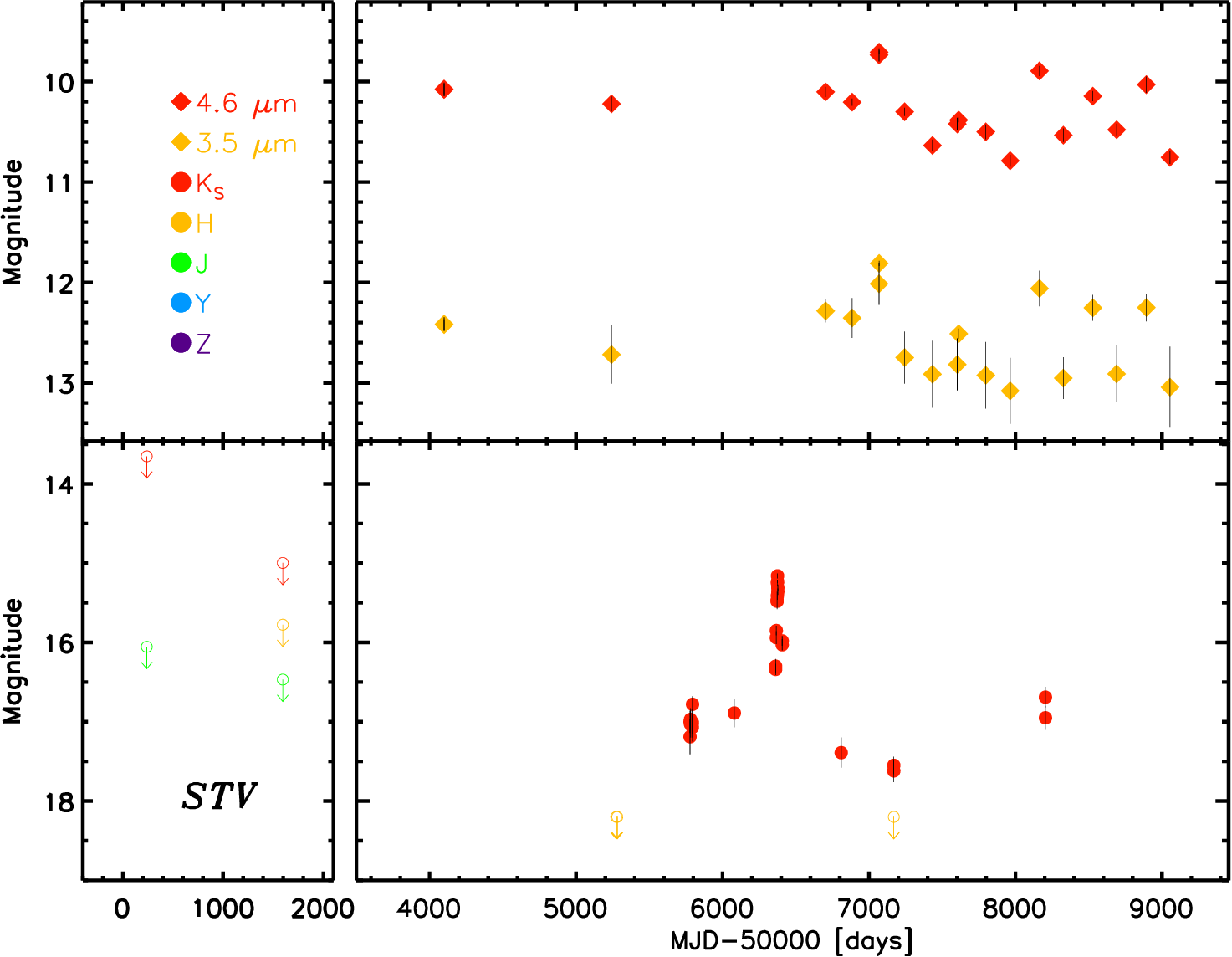}\\
\includegraphics[width=\columnwidth,angle=0]{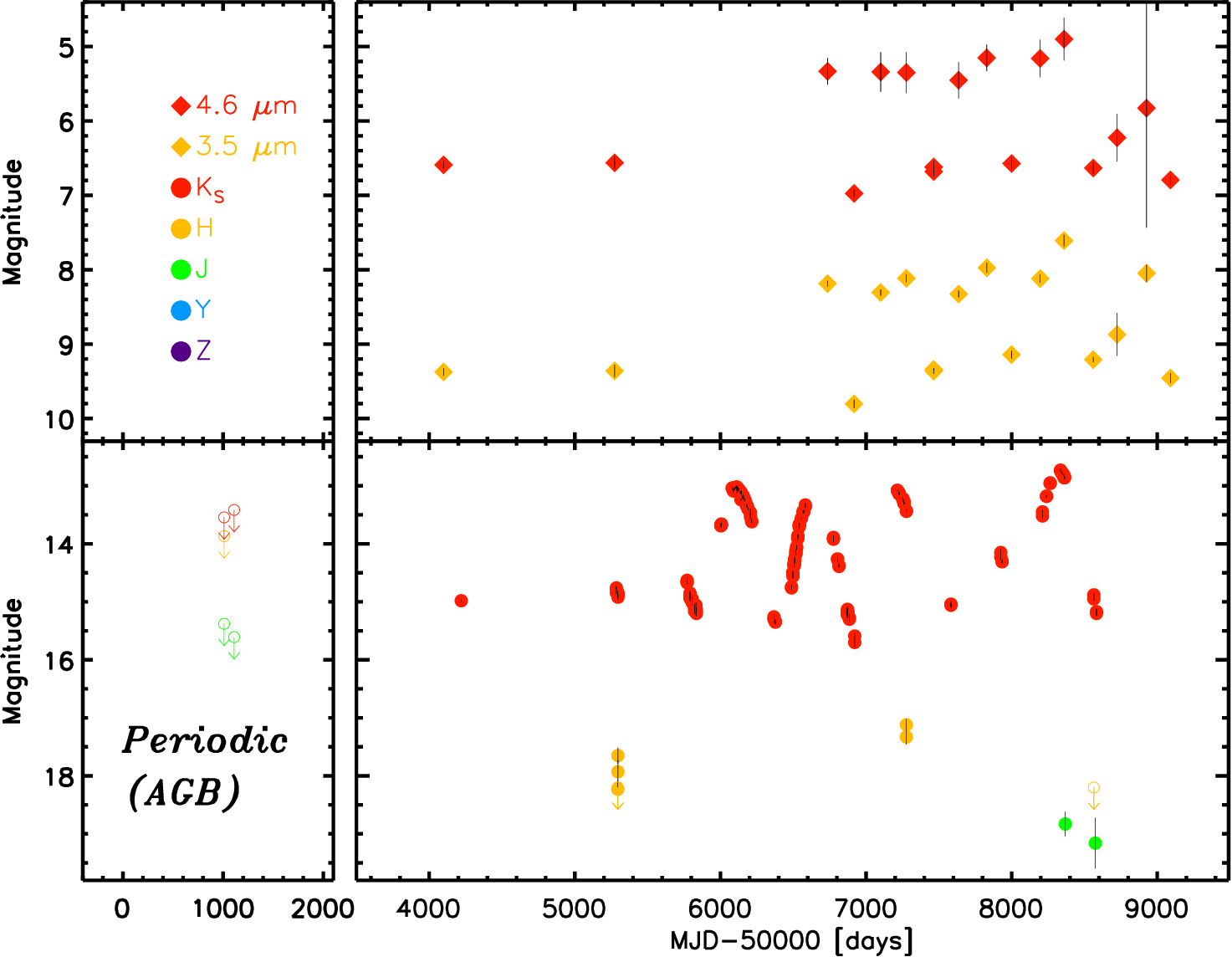}
\caption{Near- and Mid-IR light curves of five Class I YSOs that show variability that is likely not related to variable accretion. The light curves include examples of Dippers (YSO5543, top-left), Faders (YSO1069, top-right), Multiple (YSO6144, middle-left), STVs (YSO1195, middle-right), and Periodic (YSO5601, bottom).}\label{fig:yso_class}
\end{figure*}

Example of light curves that are not classified as eruptives--ccp17 are shown in Fig. \ref{fig:yso_class}. 

The majority of periodic objects show smooth sinusoidal variability. We find that objects with these light curves also show low-amplitude, nearly sinusoidal variability in the mid-IR and are classified as candidate AGB stars if we follow the criteria from \citet{2008Robitaille}. Therefore is likely that these sources are AGB stars contaminating our sample of Class I YSOs \citep[see also][]{2022Guo}. Dippers on the other hand are most likely associated with variable extinction due to structures in the accretion disc \citep{2017Contreras_a}. Interestingly, such structures could be linked to interaction with a planetary or binary companion. Observation of AA Tau \citep{2017Loomis} reveals a misaligned inner and outer disc; the long term fading likely arises from obscuration by flows crossing from the outer into the inner disc \citep[but also see][]{2021Covey}. Such misalignment could be related to planet migration \citep{2018Nealon}. In RW Aur, the observed variability might be caused by tidal perturbations from interactions with a companion \citep{2018Rodriguez} or a dusty wind \citep{2019Dodin}.

The study of other types of variability, although interesting, is beyond the scope of this paper. In the following, we will focus on the YSOs that show light curves that can be associated with episodic accretion.

\section{Accretion-driven outbursts}\label{sec:acc_out}

The initial inspection of light curves provides a rough classification to separate those YSOs with variability that is most likely due to episodic accretion. In this sense, all of the YSOs classified as {\it Eruptive--ccp17} are included in the final list of accretion-driven outbursts.

We also include objects with light curves that are originally classified as either {\it Faders}, {\it STVs} or {\it Periodic} objects, but that show high-amplitude variability in the mid-IR. Large brightness changes at these wavelengths are expected in accretion driven outbursts \citep{2022Hillenbrand}. A classification of a light curve as {\it STV} could result as a combination of the repetitive nature of EX Lupi type outbursts and the cadence of VVV observations. The interaction with a planetary or stellar companion can lead to periodic outbursts \citep[see e.g.][]{2020Lee, 2022Guo} and therefore a classification as {\it Periodic}. Finally, {\it Faders} could be related to outbursting YSOs coming back to quiescence.

We note that from this point onward we will refer as candidate eruptive YSOs to all of the objects where we believe that variability is driven by large changes in the accretion rate.

We finally find 97 Class I YSOs that can be classified as candidate eruptive YSOs based on their long-term light curves. From the list, 70 YSOs correspond to new discoveries. The remaining 27 YSOs are part of previous studies of variability in the VVV survey, with 16 of them having follow-up spectroscopic observations (\citealp{2017Contreras_a,2017Contreras,2020Guo,2021Guo}, Guo et al., submitted) and a few sources independently identified in VIRAC2 searches by Lucas et al., (submitted). In the following we discuss the overall properties of the sample. Table \ref{tab:allvar} contains the spectral indices, distance, near- and mid-IR amplitudes, IR luminosity, duration and classification of the outburst for each source. In the same table we include, when available, the designation from previous VVV studies as well as classification from spectroscopic observations.

\subsection{Outburst classification}

Variable accretion in YSOs can be caused by a variety of different physical mechanisms and there may exist a continuum of outbursting behaviour with a range of amplitudes (0.2--7 mag in the optical) and timescales \citep[0.1 d to 100 yr,][]{2017Cody}. Outbursts lasting from 0.1 days to a few months could be explained by viscous and magnetic instabilities at the boundary between the stellar magnetosphere and the accretion disc \citep{2008Kulkarni, 2012Dangelo} whilst longer duration events (a few to up to 100 years) are potentially attributed to several physical mechanisms (see Section \ref{sec:intro}).

\begin{figure}
\includegraphics[width=\columnwidth,angle=0]{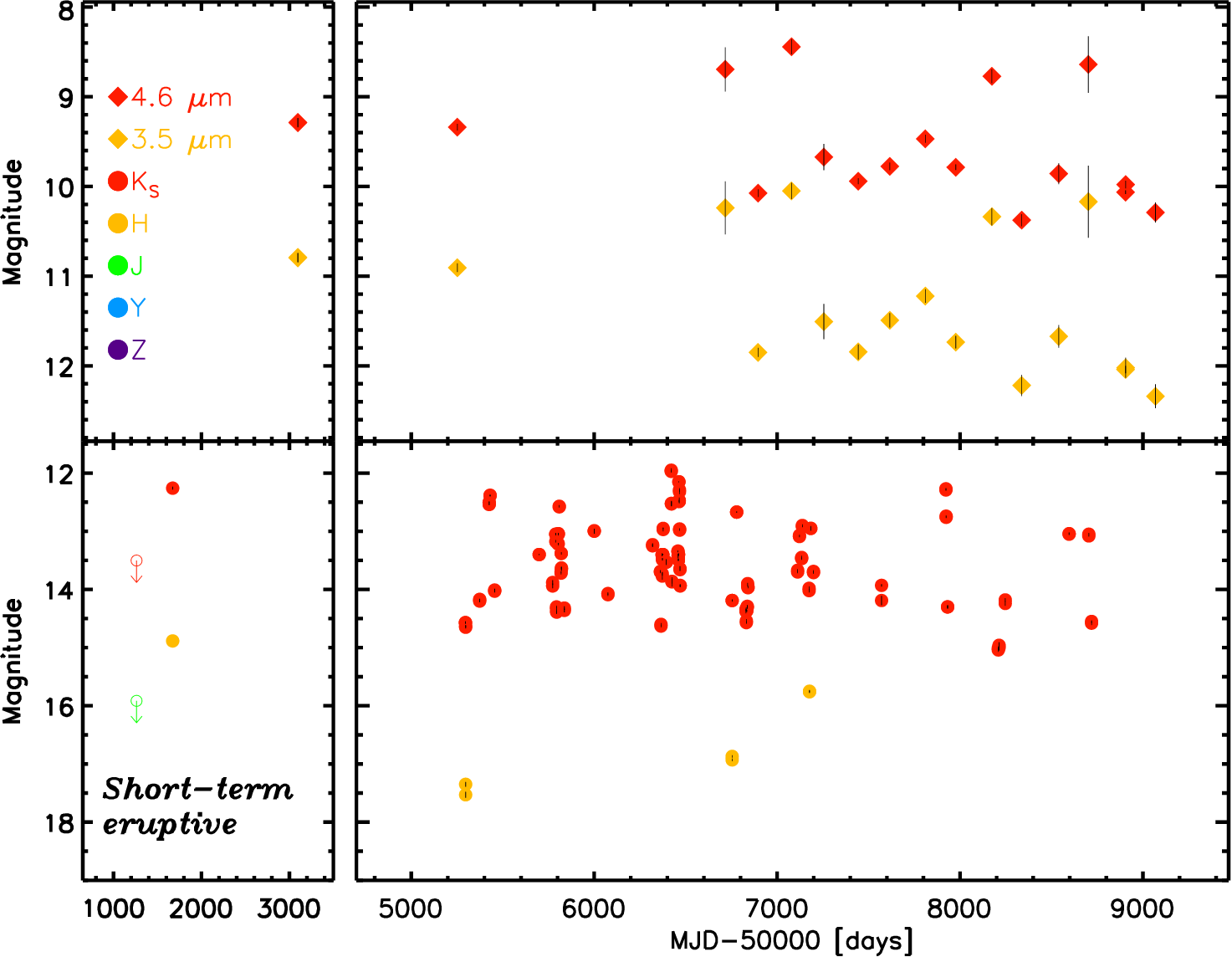}
\includegraphics[width=\columnwidth,angle=0]{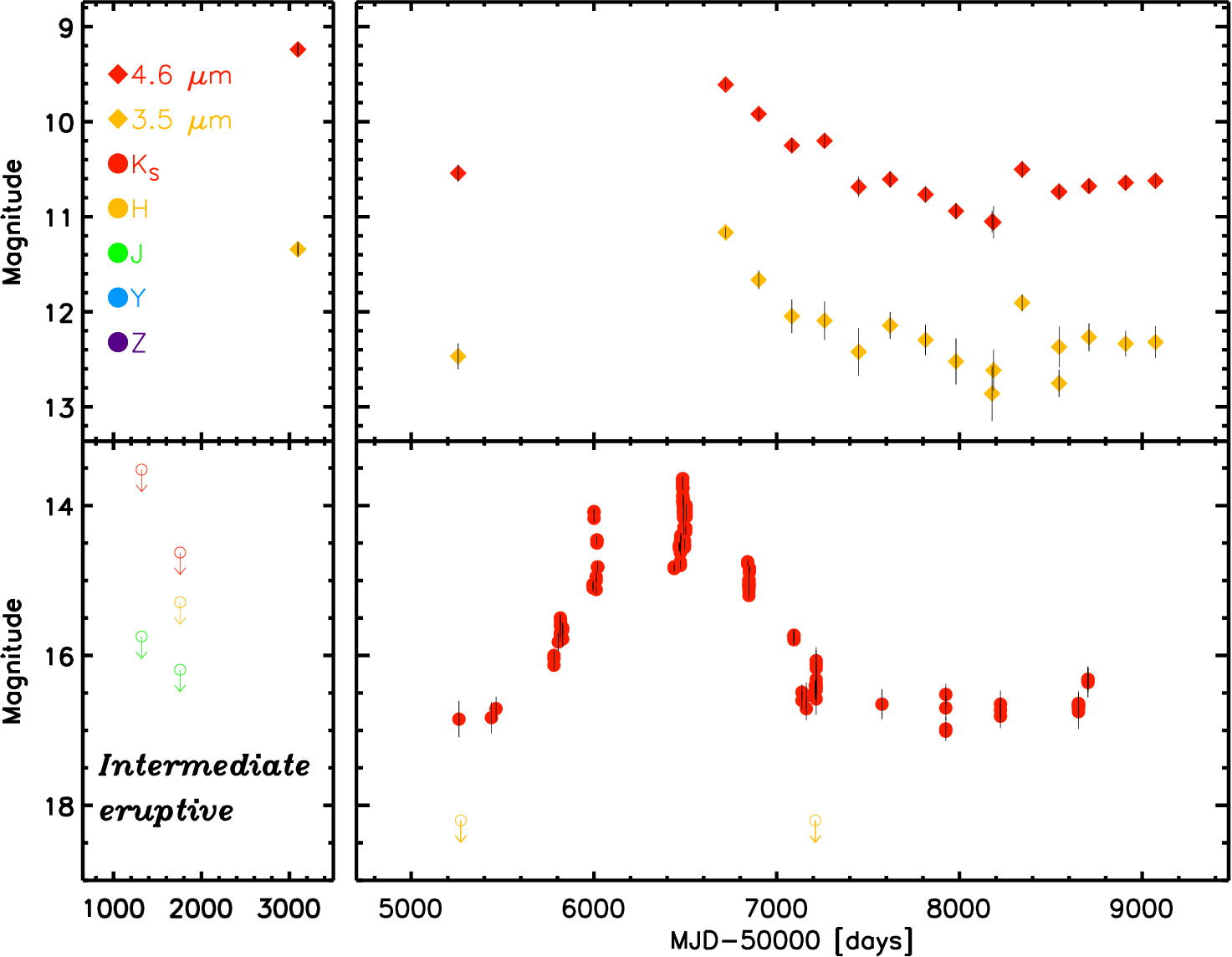}
\includegraphics[width=\columnwidth,angle=0]{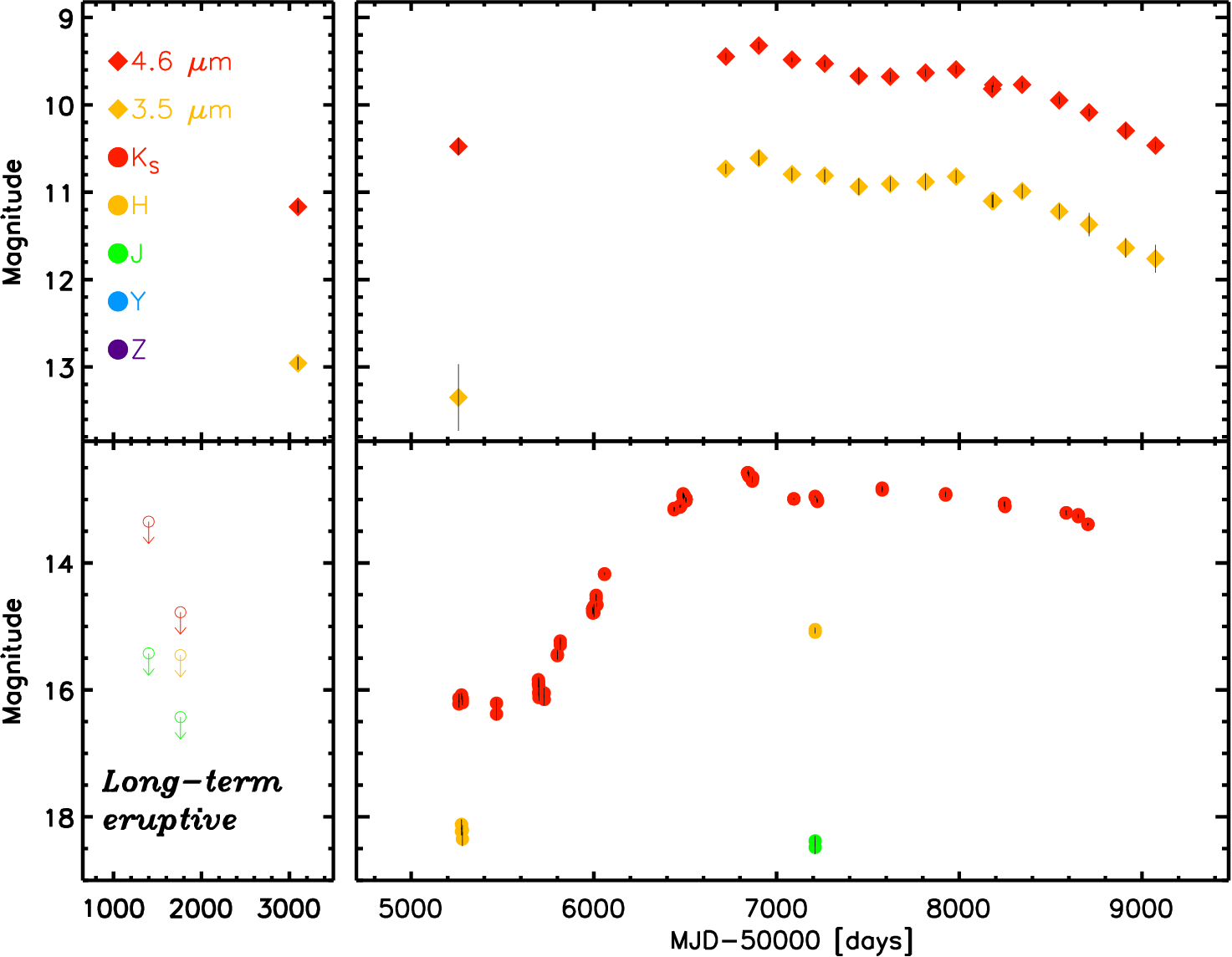}
\caption{Near- and mid-IR light curves of three eruptive YSOs. These represent examples of outbursts with short- (top, YSO5801) intermediate- (middle, YSO2822) and long-duration (bottom, YSO3017), as defined in the main text.}\label{fig:out_class}
\end{figure}

These type of events have been previously observed at optical, near- and mid-IR wavelengths. YSOs showing accretion-related variability are divided into different classes based on their photometric and spectroscopic characteristics. The short-term bursters defined by \citet{2013Findeisen, 2014Cody, 2014Stauffer} display stochastic bursts that last usually less than ~50 days and/or are of low-amplitude ($\Delta$R$<2$~mag). EX Lupi-type outbursts \citep{2012Lorenzetti}, show $\Delta t<1-2$~yr and spectra during outburst which are rich in emission lines. Long duration outbursts ($\Delta t>$10 yr) are usually classified as FUors and also show absorption dominated spectra (Na {\sc i} D, CO, H$_{2}$O) during outburst \citep{2018Connelley}. There are an increasing number of eruptive YSOs, such as V1647 Ori, ASSASN13-db, Gaia19bey and the eruptive YSOs from VVV \citep{2017Contreras, 2021Guo}, that show outburst duration that are between those of EX Lupi-type  and FUors, and show a mixture of spectroscopic characteristics between the two classes. In fact, \citet{2021Guo} shows that EX Lupi-type emission line spectra is very common even among long-duration outbursts.

\begin{figure}
	\resizebox{1\columnwidth}{!}{\includegraphics{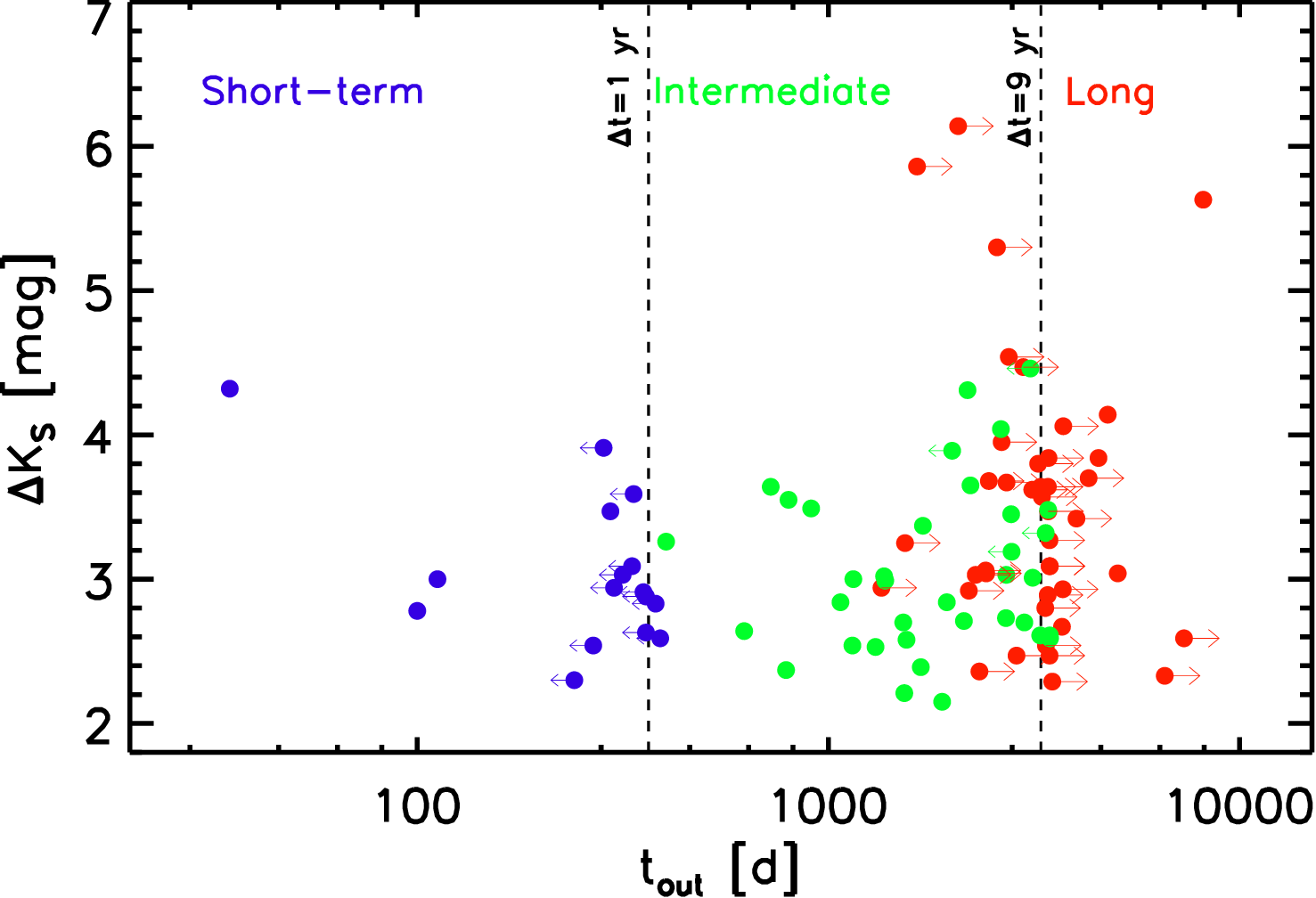}}
	 \caption{K$_{\rm S}$ amplitude, $\Delta K_{\rm S}$ versus outburst duration, $\Delta t$, for the 97 YSOs classified as eruptive in this work. Colours indicate eruptive YSOs classified as short-term (blue), intermediate (green) and long-duration (red). Dashed lines mark 1 and 9 yrs as the times used to define the different classes.}
    \label{fig:time}
\end{figure}

Because we lack spectroscopic data for the majority of the eruptive YSOs, we cannot provide a classification of our sample that strictly follows the discussed sub-classes. We simply divide YSOs according to the duration of the observed outbursts, $t_{out}$. To measure $t_{out}$ we visually inspect the long-term $K_{\rm S}$ light curve and determine the points corresponding to the start and end of the outburst. For YSOs with outbursts that are still ongoing at the latest epoch of VVV, $t_{out}$ is flagged as a lower limit. Based on the values of $t_{out}$, we divide our sample into short-term ($t_{out}\leq1$~yr), intermediate ($1<t_{out}\leq9$~yr) and long-term ($t_{out}>9$~yr) outbursts. From 97 objects, we classify 43 YSOs as long-term, 37 as intermediate and 17 as short-term outbursts.  

\begin{figure}
\resizebox{1\columnwidth}{!}{\includegraphics{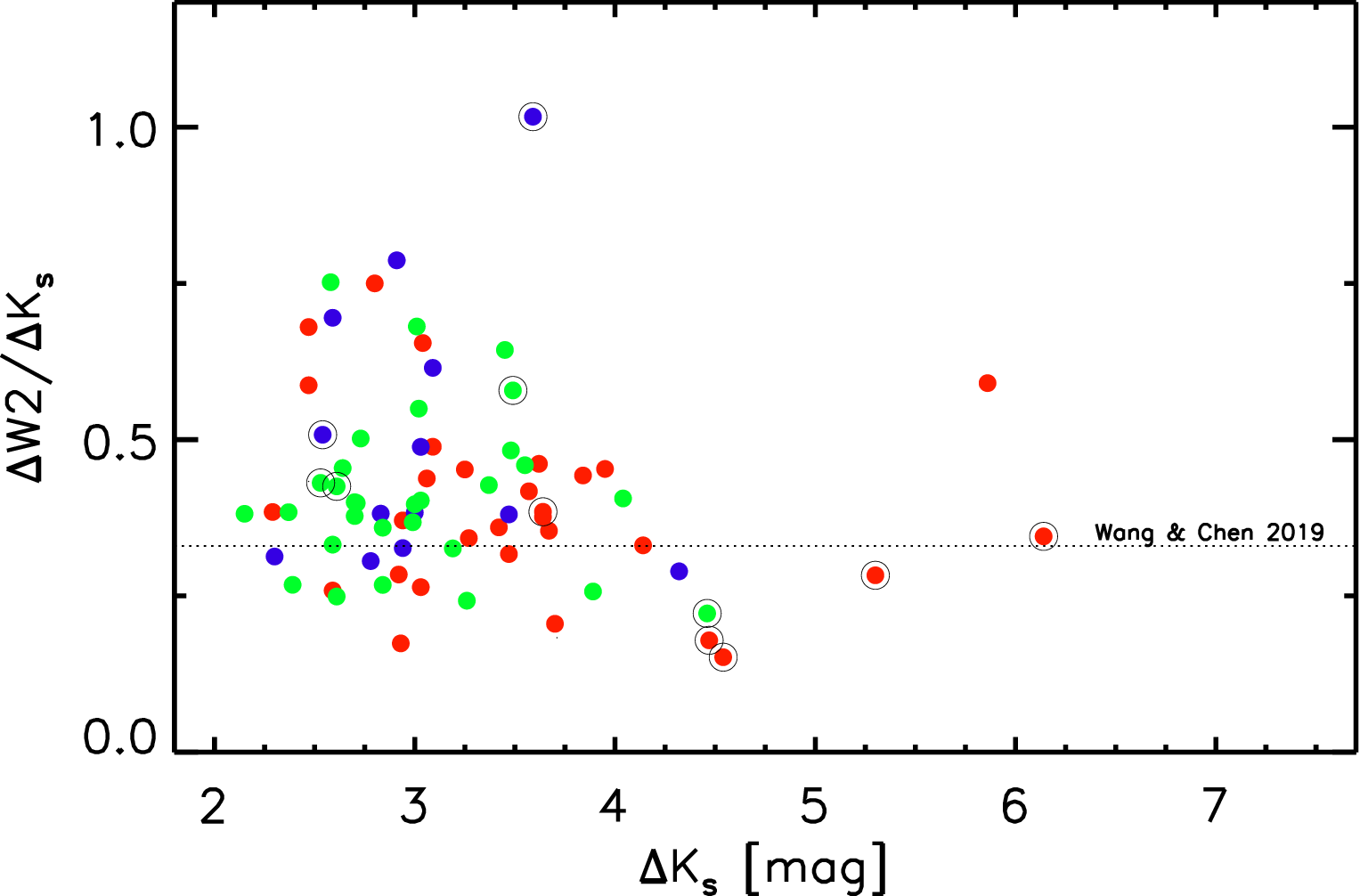}}
\caption{Ratio of mid-IR ($W2$) to near-IR ($K_{\rm S}$) amplitudes versus near-IR ($K_{\rm S}$) amplitude for YSOs classified as eruptive, and that have nearly contemporaneous observations from the VVV and {\it WISE} surveys. Black open circles mark spectroscopically confirmed eruptive YSOs. The dotted line in the plot marks the expected ratios for the \citet{2019Wang} extinction law.}
\label{fig:reddening}
\end{figure}

Figure \ref{fig:time} shows $\Delta K_{\rm S,all}$ versus $t_{out}$ for the sample of Class I eruptive YSOs, where $\Delta K_{\rm S,all}$ is the amplitude (maximum minus minimum brightness) of the long-term K$_{\rm S}$ light curve, i.e. including data from previous near-IR surveys, when available. The comparison shows that there is not a clear distinction between the different classes as they all show a similar distribution in amplitudes with $2<\Delta<4.8$~mag. The largest amplitudes ($\Delta>4.8$ mag), however, are associated exclusively with long-duration outbursts. For the 16 objects with published spectroscopic data, emission line spectra dominate the sample and are observed in short, intermediate and long-term eruptive YSOs. \citet{2021Guo} concludes that the observation of emission line spectra even in the long-duration events, which are usually associated with FUor outbursts, indicates that the magnetosphere is still in control even for these extreme events. Only 2 out of 16 YSOs with spectroscopic data show FUor-like absorption spectra, and these are found in the long duration class. More examples of very high amplitude events and their spectra will be published in a VVV-selected sample (Lucas et al., submitted; Guo et al., submitted.).

Figure \ref{fig:reddening} shows the comparison between the K$_{\rm s}$ and $W2$ amplitudes for the sample with contemporaneous observations over this wavelength range. In the figure we mark the expected values of $\Delta W2/\Delta K_{\rm s}$ if variability is due to extinction along the line of sight \citep[which lie between 0.4--0.5][]{1999Fitzpatrick, 2009Mcclure, 2019Wang}. YSOs where the variability is driven by accretion changes are expected to fall above these lines \citep{2021Guo}. Some objects in our sample fall below this boundary which could be explained by the fact that VVV and NEOWISE, although almost contemporaneous (within a few months), might still give imprecise $\Delta W2/\Delta K_{\rm s}$ values for individual YSOs. We note, however, that 9 out of 11 spectroscopically confirmed eruptive YSOs (shown by black open circles in the figure) fall below the \citet{2009Mcclure} line. Therefore, the location of YSOs in Fig. \ref{fig:reddening} does not necessarily imply that variability is being driven by changes in the extinction. However, we cannot discard that some YSOs with this type of variability might be contaminating our sample.

\subsection{Distances and accretion luminosity}

Distances are estimated from a literature search for Star Forming Regions (SFR) located within 5\arcmin~ from the YSO. Some objects that are found in previous VVV studies have distances that are determined by radial velocity measurements of emission or absorption features in their spectra \citep{2021Guo}. 

 Following \citet{2017Lucas} and \citet{2021Guo}, we want to estimate the in-outburst infrared luminosities (L$_{IR}$) of our eruptive Class I YSOs using photometry in the 2--24 $\mu$m wavelength range. The first step to measure (L$_{IR}$) was to select the mid-IR (3--24 $\mu$m) photometry either from {\it Spitzer} IRAC/MIPS \citep{2015Gutermuth} or the {\it WISE} Allsky data release \citep{2013Cutri}. These measurements can be considered as being nearly contemporaneous in the 3--24 $\mu$m range.  We then determine the difference $\Delta_{IRAC1/W1}$ and $\Delta_{IRAC2/W2}$ from the contemporaneous observations to the brightest epochs in the mid-IR light curve of the YSO. To build the in-outburst luminosities, the first two passbands in the contemporaneous mid-IR photometry are shifted by $\Delta_{IRAC1/W1}$ and $\Delta_{IRAC2/W2}$. The magnitude change in passbands beyond 4.6 $\mu$m is assumed to be $\Delta_{IRAC2/W2}$. Finally, the K-band data point is taken as the brightest measurement arising from VVV/2MASS/DENIS/UKIDSS surveys.

To determine in-outburst infrared luminosities we integrated the SEDs over the 2--24 $\mu$m wavelength range and used the distances from Table \ref{tab:allvar}. Given the uncertainties that are inherent to the process of estimating L$_{IR}$, we do not correct for extinction effects nor try to determine bolometric luminosities as these corrections would only add more uncertainty.

\begin{figure}
\flushright
\includegraphics[width=\columnwidth,angle=0]{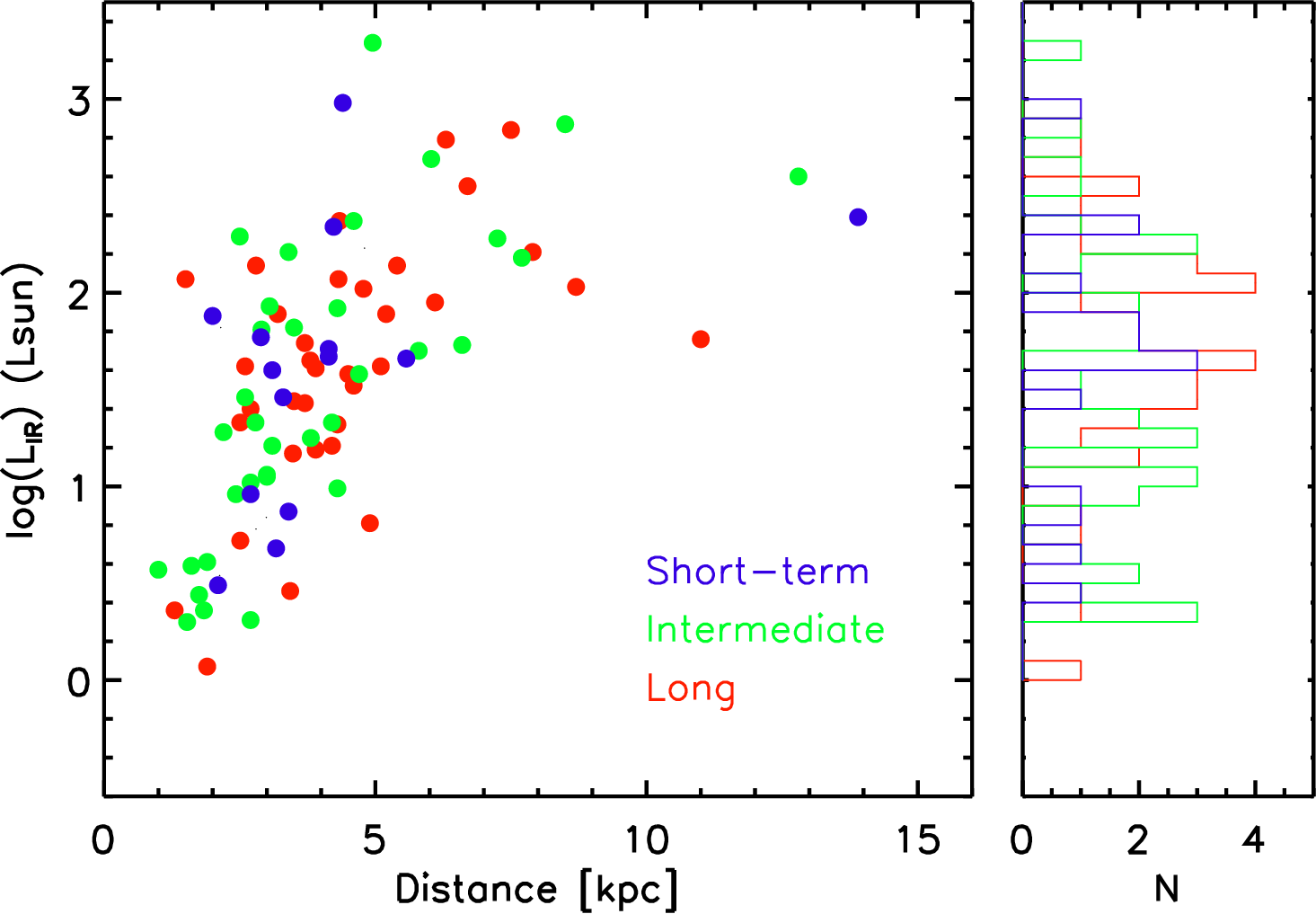}
\caption{IR luminosity versus distance for YSOs classified as eruptive in this work.}
\label{fig:luminosity}
\end{figure}

Figure \ref{fig:luminosity} shows the in outburst luminosities versus distance for the sample where we are able to measure these parameters. The results appear to show that long-term outbursts are preferentially located at higher luminosities, however, these high luminosities can be reached even for short-duration outbursts. The results of table 5 in \citet{2021Guo}, also show that short-duration outbursts can reach high luminosities.

\begin{table*}
	\centering
	\caption{The 97  candidate eruptive variable stars detected in our analysis.}
	\label{tab:allvar}
\resizebox{0.93\textwidth}{!}{
   	\begin{tabular}{lccccccccccccc} 
		\hline
		ID & SIMBAD ID &$\alpha_{24}$ & LC class &  $\Delta K_{\rm S}$ & $\Delta W1$ & $\Delta W2$ & $\Delta$t & $\log L_{IR}$ & distance & distance (ref) & separation & VVV ID & Spec type \\
		\hline
            YSO5558 & SSTGLMC G000.5969$-$00.8711   &   0.8 & long-term &  6.1 &  3.5 &  2.4 & $>$ 2068 &  2.6 &  3.5$^{\dagger}$ & -- & -- & L222\_167  & emission \\
            YSO5572 & [RMB2008] G002.1471$+$00.2834 &   0.6 & intermediate &  4.3 &  0.9 &  1.2 &  2180 &  1.0 &  2.4 & [PLW2012] G002.147$+$00.274+004.0 & 34\arcsec & L222\_164 & -- \\
            YSO5694 &     --   &  -0.3 & intermediate &  2.5 &  1.2 &  1.1 &  1302 &  0.6 &  1.9 & VVVv32 & source & VVVv32  & emission  \\
            YSO5801 & 2MASS J15401442$-$5440034     &  $-$0.3 & short-term &  3.1 &  2.1 &  1.9 &   333$<$ &  1.5 &  3.3 & [PLW2012] G325.853$+$00.495$-$060.6 & 259\arcsec & --     & -- \\
            YSO5827 & 2MASS J15575037$-$5357334     &   0.3 & intermediate &  4.5 &  1.8 &  1.0 &  3100$<$ &  1.9 &  3.0 & [GBN2014] 33 & 52\arcsec & L222\_32     & outflow \\
            YSO5844 & 2MASS J15142364$-$5802344     &  $-$0.2 & long-term &  3.8 &  1.3 &  1.7 & $>$3431 &  1.6 &  5.1 & Stim5 & source & Stim5   & emission  \\
            YSO5870 &            --              &  $-$0.2 & short-term &  2.3 &  1.5 &  1.0 &   241$<$ &  1.0 &  2.7 & AGAL G333.099$-$00.094 & 67\arcsec & VVVv252 & -- \\
            YSO5921 & VVVv717&  $-$0.3 & long-term &  4.5 &  2.1 &  0.8 & $>$2979 &  2.0 &  6.1 & VVVv717 & source & VVVv717 & absorption \\
            YSO5956 & SSTGLMC G341.8798$+$00.3850   &  $-$0.1 & long-term &  3.7 &  2.0 &  1.0 & $>$2456 &  1.2 &  3.9 & DR4\_V44 & source & DR4\_V44 & emission  \\
            YSO6020 & 2MASS J17180413$-$3852225     &   0.7 & short-term &  2.8 &  2.6 &  2.0 &   380$<$ &  1.9 &  3.5$^{\dagger}$ & -- & -- & --  & -- \\
            YSO6029 &  -- &  $-$0.3 & intermediate &  2.7 &  1.1 &  1.1 &  2136 &  1.3 &  2.2 & HIGALBM350.3555$-$0.9291 & 3.2\arcsec  & --  & -- \\
		YSO5413 & SSTGLMC G007.7524$-$00.8625 & 0.6 & long-term &  3.04 & 4.7 & 3.3 & $>2420$ & 2.37 & 4.34 & [MML2017] 1106 & 315\arcsec & -- & -- \\
            YSO6035 & [RMB2008] G350.6642$+$00.8936 &  $-$0.1 & intermediate &  4.0 &  1.2 &  1.6 &  2628 &  0.6 &  1.6 & [WMA2013] J171753.082$-$361459.51 & 103\arcsec & L222\_77     & -- \\
YSO6078 & SSTGLMC G353.4039$-$00.0664   &  $-$0.3 & short-term &  4.3 &  1.9 &  1.7 &    35 &  1.7 &  5.6 & AGAL G353.396$-$00.071 & 32\arcsec & --  & -- \\
YSO6088 & [RMB2008] G353.8229$+$00.2585 &   1.1 & intermediate &  3.6 &  2.4 &  2.6 &   724 &  2.2 &  3.4 & DR4\_V55 & source & DR4\_V55 & outflow \\
YSO6151 & [RMB2008] G357.4063$-$00.5578 &  $-$0.3 & long-term &  5.3 &  2.5 &  1.5 & $>$2571 &  1.6 &  4.5 & DR4\_V67 & source & DR4\_V67 & outflow \\
YSO6166 & SSTGLMC G358.5446$-$00.3383   &   0.4 & intermediate &  3.9 &  2.3 &  1.3 &  2000$<$ &  2.9 &  8.5 & [PDT2018] G358.545$-$00.338 & 1.2\arcsec & -- & -- \\
YSO6198 & SSTGLMC G359.7288$-$00.4063   &   0.8 & long-term &  4.5 &  2.0 &  1.0 & $>$2747 &  2.1 &  1.5 & [KC97c] G359.7$-$00.4 & 4.11\arcsec & L222\_148     & emission \\
YSO227  & 2MASS J12212799$-$6301040     &   0.3 & short-term &  2.9 &  1.2 &  1.0 &   301$<$ &  1.7 &  4.1 & [BYF2011] 208b & 99\arcsec & --     & -- \\
YSO229  & --                         &   0.7 & short-term &  2.6 &  3.5 &  3.3 &   360$<$ &  1.7 &  4.1 & [BYF2011] 208b & 32\arcsec & --     & -- \\
YSO288  & SSTGLMC G300.4749$+$00.1243   &   0.4 & short-term &  2.6 &  3.0 &  2.6 &   390$<$ &  2.3 &  4.2 & HIGALBM300.4745$+$0.1240 & 5\arcsec & --     & -- \\
YSO323  &            --              &   0.7 & intermediate &  2.6 &  1.4 &  1.9 &  1549 &  1.9 &  4.3 & AGAL G300.951$+$00.894 & 44\arcsec & --     & -- \\
YSO386  & --                         &   0.1 & intermediate &  3.0 &  1.6 &  1.2 &  2708 &  0.4 &  1.8 & [RC2004] G301.8$+$1.0$-$25.9 & 252\arcsec & --     & -- \\
YSO443  & --                         &   0.5 & long-term &  3.0 &  2.0 &  1.2 & $>$ 2282 &  1.8 & 11.0 & [KC97c] G302.5$-$00.7 & 138\arcsec & --     & -- \\
YSO492  &            --              &  $-$0.1 & long-term &  3.0 &  3.1 &  2.8 &  5056 &  0.8 &  3.0 & HIGALBM303.0625$+$0.1649 & 78\arcsec & --     & -- \\
YSO507  &            --              &   0.0 & intermediate &  2.4 &  0.9 &  0.9 &   789 &  0.4 &  1.8 & [MML2017] 4504 & 417\arcsec & --     & -- \\
YSO532  & 2MASS J12574420$-$6215063     &   0.2 & long-term &  2.5 &  2.2 &  1.7 & $>$3451 &  1.5 &  3.5$^{\dagger}$ & -- & -- & --     & -- \\
YSO577  & 2MASS J13011322$-$6225279     &  $-$0.2 & intermediate &  2.7 &  1.5 &  1.1 &  2996 &  1.3 &  4.2 & AGAL G304.089$+$00.421 & 92\arcsec & VVVv467 & -- \\
YSO649  & --                         &   0.4 & long-term &  3.2 &  1.6 &  1.6 & $>$ 1536 &  0.5 &  2.1 & [MML2017] 2754 & 165\arcsec & --     & -- \\
YSO705  & SSTGLMC G305.6061$+$00.0733   &   0.3 & long-term &  3.7 &  2.8 &  1.3 & $>$2712 &  1.6 &  3.9 & AGAL G305.581$+$00.037 & 159\arcsec & --     & -- \\
YSO751  & --                         &   0.3 & long-term &  2.4 & -- & -- & $>$ 3432 & -0.41(f) &  2.1 & AGAL G306.503$+$00.072 & 403\arcsec & --     & -- \\
YSO795  & --                         &  $-$0.1 & short-term &  2.5 &  0.9 &  1.3 &   268$<$ &  2.4 & 13.9 & VVVv53 & source & VVVv53  & emission  \\
YSO829  &            --              &   0.0 & long-term &  3.1 &  1.8 &  1.4 & $>$ 2413 &  2.8 &  7.5 & HIGALBM307.6300$-$0.2732 & 0.3\arcsec & --     & -- \\
YSO966  & 2MASS J13470753$-$6246332     &  $-$0.1 & long-term &  3.4 &  1.8 &  1.3 & $>$4013 &  1.4 &  3.5 & AGAL G309.271$-$00.552 & 152\arcsec & --     & -- \\
YSO992  &            --              &   0.1 & intermediate &  2.6 &  2.4 &  1.6 & $>$3460 &  2.2 &  7.7 & HIGALBM309.5305$-$0.6368 & 1.7\arcsec & --     & -- \\
YSO1027 &            --              &  $-$0.2 & long-term &  2.3 &  1.4 &  0.9 & $>$3504 &  2.8 &  6.3 & HIGALBM310.1512$-$0.1348 & 2\arcsec  & --     & -- \\
YSO1071 & SSTGLMC G310.9515$+$00.9553   &   0.3 & intermediate &  2.2 &  2.4 &  1.4 & $>$1891 &  1.0 &  4.3 & [MML2017] 591 & 60\arcsec & --     & -- \\
YSO1111 & SSTGLMC G311.2960$+$00.7965   &   0.4 & intermediate &  3.5 &  2.3 &  1.6 & $>$800 &  2.3 &  7.2 & HIGALBM311.2959$+$0.7966 & 1.4\arcsec & --     & -- \\
YSO1329 & SSTGLMC G312.8542$-$00.4687   &   0.4 & long-term &  2.4 &  1.6 &  1.7 & $>$ 2330 &  1.4 &  3.7 & [MML2017] 740 & 232\arcsec & --     & -- \\
YSO1473 & --                         &   0 & long-term &  3.3 &  1.4 &  1.4 & $>$ 3450 &  0.5 &  3.4 & [PLW2012] G314.326$-$00.059$-$051.2 & 333\arcsec & -- & -- \\
YSO1648 & 2MASS J14503936$-$5922540     &   0.4 & intermediate &  2.6 &  1.1 &  0.7 &  3275 &  1.8 &  2.9 & AGAL G317.596$+$00.052 & 2.1\arcsec & --     & -- \\
YSO1654 & 2MASS J14512100$-$6000276     &   0.7 & short-term &  3.6 &  3.4 &  3.5 &   336$<$ &  1.5 &  2.9 & [MML2017] 610 & 195\arcsec & VVVv118 & emission  \\
YSO1745 & --    &   0.4 & intermediate &  2.6 &  1.4 &  1.1 &  3456 &  1.5 &  2.6 & [PLW2012] G318.574$-$00.189$-$046.4 & 82\arcsec& VVVv128 & emission  \\
YSO1852 &            --              &  $-$0.1 & long-term &  3.6 &  2.2 &  2.3 & $>$ 3419 &  1.3 &  4.3 & DR4\_V15 & source & DR4\_V15 & outflow \\
YSO1935 & SSTGLMC G321.2967$+$00.9440   &   0.5 & intermediate &  3.0 &  2.3 &  1.6 &  1366 &  0.3 &  1.5 & [MML2017] 3488 & 597\arcsec & --     & -- \\
YSO1956 & --                         &   0.3 & long-term &  2.9 &  2.8 &  1.4 & $>$ 3717 &  1.4 &  2.7 & AGAL G320.416$-$00.937 & 55\arcsec & VVVv140 & -- \\
YSO1961 & 2MASS J15142364$-$5802344     &  $-$0.2 & long-term &  3.1 &  2.1 &  1.5 & $>$ 3456 &  1.3 &  2.5 & [PLW2012] G320.886$-$00.385$-$044.8 & 496\arcsec & --     & -- \\
YSO2009 &            --              &   0.6 & long-term &  2.9 &  0.8 &  1.2 & $>$ 3416 &  0.1 &  1.9 & AGAL G322.459$+$00.971 & 47\arcsec & --     & -- \\
YSO2099 & SSTGLMC G323.5111$+$00.0510 &   0.2 & long-term &  5.6 &  3.2 &  1.8 &  8161 &  1.7 &  3.7 & [PLW2012] G323.508$+$00.050$-$067.4 & 12\arcsec & --     & -- \\
YSO2345 &            --              &   0.1 & intermediate &  2.7 &  1.1 &  1.3 &  1521 &  1.2 &  3.8 & HIGALBM326.3262$-$0.2206 & 2.5\arcsec& --     & -- \\
YSO2355 & --                         &  $-$0.1 & long-term &  2.7 &  2.5 &  2.0 &  3700 &  0.8 &  4.9 & AGAL G327.109$+$00.539 & 224\arcsec & --     & -- \\
YSO2482 & --                         &   0.0 & intermediate &  3.0 &  1.4 &  1.2 &  1150 &  1.7 &  5.8 & AGAL G328.318$+$00.396 & 7\arcsec  & --     & -- \\
YSO2589 & --                         &   0.6 & intermediate &  3.3 &  2.6 &  1.4 &  3377$<$ &  1.1 &  3.0 & AGAL G328.614$-$00.466 & 278\arcsec & VVVv217 & -- \\
YSO2655 & SSTGLMC G329.7165$+$00.2135   &  $-$0.3 & long-term &  3.5 &  2.2 &  1.1 & $>$ 3426 &  1.9 &  5.2 & AGAL G329.682$+$00.144 & 280\arcsec & --     & -- \\
YSO2668 & SSTGLMC G329.1714$-$00.6045   &   0.3 & intermediate &  3.5 &  2.7 &  2.2 &  2783 &  2.4 &  4.6 & AGAL G329.111$-$00.491 & 463\arcsec & --     & -- \\
YSO2822 & SSTGLMC G331.6913$-$00.1390   &  $-$0.3 & intermediate &  3.4 &  1.7 &  1.5 &  1698 &  1.0 &  3.0 & AGAL G331.682$-$00.124 & 64\arcsec  & --     & -- \\
YSO2832 & SSTGLMC G332.6736$+$00.7864   &   0.0 & long-term &  3.6 &  2.0 &  1.4 & $>$ 3283 &  2.0 &  4.8 & [MML2017] 1248 & 382\arcsec  & --     & -- \\
YSO2876 & SSTGLMC G332.3559$-$00.1023   &   0.2 & intermediate &  2.7 &  3.9 &  1.5 &  2703 &  1.2 &  3.1 & AGAL 332.352$-$00.116 & 52\arcsec & --     & -- \\
YSO2945 & 2MASS J16203727$-$5045071     &  $-$0.3 & intermediate &  2.8 &  1.4 &  0.8 &  1941 &  1.8 &  3.5 & AGAL332.986$-$00.489 & 72\arcsec  & --     & -- \\
YSO2971 & 2MASS J16214416$-$5020415     &   0.4 & long-term &  2.3 &  0.9 &  0.7 & $>$ 6578 &  1.9 &  3.2 & AGAL G333.383$-$00.334 & 14\arcsec & VVVv263 & -- \\
YSO2972 & SSTGLMC G333.3638$-$00.3573   &   0.3 & short-term &  3.8 & -- & -- &   232 & 1.2(f) &  3.2 & AGAL G333.338$-$00.361 & 94\arcsec  & --     & -- \\
YSO3017 & VVVv270                     &   0.1 & long-term &  3.8 &  2.7 &  1.1 & $>$ 3237 &  1.5 &  4.6 & VVVv270 & source & VVVv270 & emission  \\
YSO3122 & --                         &   0.1 & short-term &  3.9 &  1.8 &  1.9 &   284$<$ &  0.9 &  3.4 & AGAL G335.284$-$00.132 & 24\arcsec & --     & -- \\
YSO3135 & 2MASS J16293653$-$4910585     &  $-$0.1 & intermediate &  3.7 &  2.0 &  0.9 &  2216 &  1.0 &  2.7 & AGAL G335.099$-$00.429 & 31\arcsec & --     & -- \\
YSO3250 & SSTGLMC G337.1347$-$00.1374   &  $-$0.2 & intermediate &  2.5 &  2.1 &  1.1 &  1144 &  1.6 &  4.7 & AGAL G337.141$-$00.152 & 56\arcsec & --     & -- \\
YSO3277 & SSTGLMC G337.0688$-$00.4182   &   2.4 & short-term &  3.5 &  2.4 &  2.4 &   295 &  3.0 &  4.4 & PGCC G337.13$-$00.40 & 248\arcsec  & --     & -- \\
YSO3310 & SSTGLMC G337.4423$-$00.4058   &   0.8 & short-term &  3.0 &  3.1 &  1.6 &   112 &  1.6 &  3.1 & AGAL G337.451$-$00.382 & 91\arcsec  & --     & -- \\
YSO3328 & 2MASS J16394877$-$4548480     &   0.3 & long-term &  3.1 &  2.2 &  1.4 & $>$ 3452 &  2.1 &  4.9 & GAL 338.74$+$00.64 & 285\arcsec  & VVVv721 & absorption \\
YSO3350 & SSTGLMC G337.5495$-$00.6526   &   0.1 & intermediate &  3.2 &  1.6 &  1.0 &  2790$<$ &  2.6 & 12.8 & HIGALBM337.5352$-$0.5995 & 204\arcsec & --    & -- \\
YSO3382 & SSTGLMC G338.4541$-$00.1540   &   0.1 & long-term &  4.1 &  2.2 &  1.4 & $>$ 3727 &  1.7 &  3.7 & AGAL G338.429$-$00.209 & 217\arcsec & --     & -- \\
YSO3416 & 2MASS J16434470$-$4645386     &  $-$0.2 & long-term &  2.8 &  2.2 &  2.0 & $>$ 3370 &  2.1 &  2.8 & AGAL338.484$-$00.566 & 14\arcsec & --     & -- \\
YSO3539 & SSTGLMC G340.1158$-$00.3678 &  $-$0.0 & long-term &  3.6 &  1.3 &  1.5 & $>$ 3306 &  1.2 &  3.5 & [AAL2018] G340.106$-$00.354 & 61\arcsec & --     & -- \\
YSO3565 & SSTGLMC G341.2834$+$00.3193   &   0.1 & intermediate &  3.5 &  2.9 &  2.0 &   908 &  1.7 &  6.6 & DR4\_V42 & source & DR4\_V42 & emission  \\
YSO3632 & SSTGLMC G341.9204$+$00.1989   &   0.2 & long-term &  4.0 &  2.6 &  1.7 & $>$ 2638 &  2.2 &  7.9 & AGAL G341.907$+$00.202 & 46\arcsec & --     & -- \\
YSO3770 & --                         &   0.2 & intermediate &  3.5 &  3.0 &  1.8 & $>$ 3423 &  1.3 &  2.8 & AGAL G343.422$-$00.331 & 29\arcsec  & --    & -- \\
YSO3771 &            --              &   1.0 & intermediate &  2.2 &  1.1 &  0.9 &  1530 & -- &  2.5 & AGAL G343.422$-$00.331 & 9\arcsec & --     & -- \\
YSO3775 & [CLM2017] VVVv367           &   1.1 & intermediate &  3.0 &  2.4 &  1.6 & $>$ 1375 &  2.3 &  2.5 & AGAL342.584$-$01.011 & 31\arcsec & VVVv367 & -- \\
YSO3793 & 2MASS J17005692$-$4256374     &  $-$0.1 & long-term &  3.7 &  2.2 &  1.0 & $>$ 4295 &  2.1 &  5.4 & [MML2017] 935 & 100\arcsec & VVVv381 & -- \\
YSO3911 &            --              &   0.4 & short-term &  2.9 &  3.2 &  2.3 &   355$<$ &  1.9 &  2.0 & HIGALBM345.2841$-$0.1534 & 1.3\arcsec & --     & -- \\
YSO3948 & 2MASS J17072449$-$4109075     &  $-$0.1 & long-term &  2.9 &  1.2 &  0.9 & $>$ 2196 &  1.6 &  2.6 & AGAL G345.521$-$00.356 & 2.61\arcsec & --     & -- \\
YSO4009 & SSTGLMC G348.1306$+$00.5135   &   0.1 & intermediate &  2.6 &  2.4 &  1.2 &   624 &  0.6 &  1.0 & AGAL G348.156$+$00.506 & 95\arcsec & --     & -- \\
YSO4048 & SSTGLMC G349.0052$+$00.6005   &   0.7 & intermediate &  3.3 &  1.2 &  0.8 & $>$  403 &  2.7 &  6.0 & [MML2017] 5723 & 252\arcsec & VVVv806 & -- \\
YSO4098 &            --              &   0.2 & intermediate &  2.4 &  0.8 &  0.7 &  1678 &  0.8 &  2.8 & AGAL G349.459$+$00.166 & 15\arcsec & VVVv809 & -- \\
YSO4195 & --                         &   0.4 & intermediate &  3.0 &  1.7 &  1.6 &  3139 &  3.3 &  4.9 & [MML2017] 2710 & 311\arcsec & --     & -- \\
YSO4209 &            --              &  $-$0.2 & intermediate &  2.8 &  1.5 &  0.9 &  1070 &  0.3 &  2.7 & AGAL G350.017$-$00.519 & 192\arcsec  & --     & -- \\
YSO4322 & --                         &  $-$0.2 & long-term &  2.5 &  2.0 &  1.3 & $>$ 3381 &  2.1 & 17.3 & AGAL G352.872$+$00.289 & 289\arcsec & --     & -- \\
YSO4331 &            --              &  $-$0.0 & short-term &  2.8 &  1.3 &  0.8 &   100 &  0.7 &  3.2 & HIGALBM351.5317$-$0.5742 & 11\arcsec  & --     & -- \\
YSO4339 &                             &  $-$0.3 & long-term &  2.6 &  2.8 &  1.1 & $>$ 7327 &  2.0 &  8.7 & AGAL352.226$-$00.171 & 90\arcsec & --     & -- \\
YSO4450 & --                         &  $-$0.1 & short-term &  2.9 &  1.9 &  2.1 &   360$<$ &  0.5 &  2.1 & AGAL G352.142$-$01.016 & 21\arcsec  & --     & -- \\
YSO4655 & SSTGLMC G356.2147$-$00.2903   &   0.1 & intermediate &  2.7 & -- & -- &  3100$<$ & 0.56(f) &  3.5$^{\dagger}$ & -- & -- & --     & -- \\
YSO4885 & 2MASS J17472421$-$2923016     &  $-$0.0 & long-term &  3.7 & -- & -- &  6918 & 1.09(f) &  3.5$^{\dagger}$ & -- & -- & --   & -- \\
YSO4971 & SSTGLMC G000.5619$-$00.8825   &   0.9 & long-term &  3.6 &  2.9 &  1.8 & $>$ 3132 &  2.4 &  3.5$^{\dagger}$ & -- & -- & --     & -- \\
YSO5298 &            --              &   0.4 & long-term &  2.5 &  2.3 &  1.4 & $>$ 2868 &  1.6 &  3.8 & AGAL G006.339$-$00.746 & 63\arcsec  & --     & -- \\
YSO5385 &            --              &  $-$0.2 & long-term &  2.9 &  2.7 &  1.8 & $>$ 1346 &  1.2 &  4.2 & AGAL007.999$-$00.506 & 210\arcsec & --     & -- \\
YSO6823 &            --              &   0.4 & long-term &  3.8 &  2.7 &  2.3 &  4539 &  0.7 &  2.5 & HIGALBM303.5020$-$1.2318 & 5\arcsec & --     & -- \\
YSO6940 &            --              &   0.1 & long-term &  4.1 &  2.6 &  1.4 &  4780 &  0.4 &  1.3 & AGAL G352.609$-$01.097 & 189\arcsec  & --     & -- \\
YSO6652 & --                         &   0.6 & short-term &  3.0 &  0.8 &  1.5 &   316$<$ &  2.1 &  3.5$^{\dagger}$ & -- & -- & --     & -- \\
		\hline
		\multicolumn{14}{l}{$\dagger$ For these YSOs no distance information could be found in the literature. In these cases we assume a distance of 3.5 kpc}\\
	\end{tabular}}
\end{table*}

\section{Incidence of episodic accretion}\label{sec:inc_out}

In the study of the possible effects of accretion outbursts in star and planet formation, usually only the most extreme, long-lasting events (a.k.a FUors) are considered. However, shorter duration events might also have an impact, for example, \citet{2023Wang} estimate that EX Lupi accretes about 0.1 Earth masses during a large outburst, and that in general the mass accreted during outburst is two times the mass accreted during quiescent phases. EX Lupi-type outbursts may also contribute to the build-up of the crystalline dust component ubiquitously seen in comets \citep{2019Abraham, 2023Kospal}. 

Prior to estimating the incidence of episodic accretion from the number of outbursts detected in our sample, we need to take into consideration the effects on this numbers from contamination of non-YSOs, from other classes of variable stars, our choice of amplitude cut, and the definition of the Class I stage.

\subsection{Selection effects}

The selected amplitude cut to search for eruptive YSOs probably has an effect on the observed proportion of objects with different outbursts duration. For example, the VVV spectroscopic defined sample of \citet{2021Guo}, which is mostly comprised of YSOs with outbursts duration of less than 9 yr, shows 25 out of 55 objects with $\Delta K_{\rm S}<2$~mag. This could explain the larger proportion of objects with long-term outbursts found in our work. In addition, isolated short-term outbursts might not be detected due to the cadence of VVV observations.

To study the effects of amplitude and sampling we built synthetic light curves of outbursts assuming a wide range of outbursts duration, $\Delta t_{out}=$30--4200 d, and interval between outbursts, $\Delta t_{int}=$90--12000 d, with the intervals always being larger than the duration of the outbursts. The light curves are generated for two different amplitude ranges, $\Delta K=1.5$--3 mag and $\Delta K=3$--5 mag (see Appendix \ref{app:a} for the details of how the synthetic light curves are generated).

\begin{figure}
	\resizebox{\columnwidth}{!}{\includegraphics{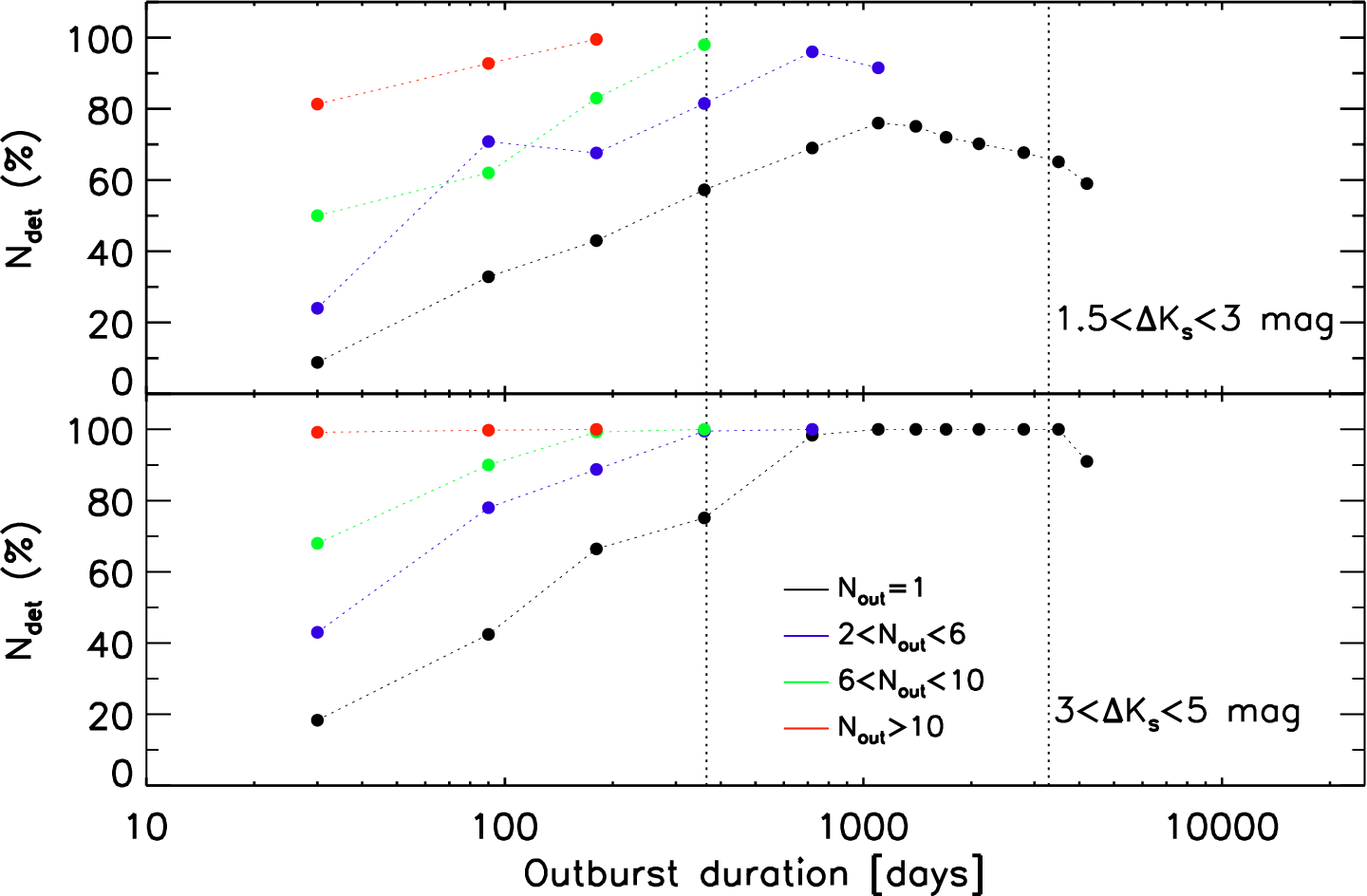}}
	 \caption{Typical number of synthetic outbursts detected by the VVV sampling}
    \label{fig:ndet}
\end{figure}

The modelling of light curves shows that the probability of detecting outbursts with the VVV sampling strongly depends on the outburst duration, amplitude and whether these repeat during the coverage of VVV observations. Figure \ref{fig:ndet} shows the number of times (in percentage) the YSO would be detected as a high-amplitude ($\Delta K_{\rm s}>2$~mag) variable star vs the outburst duration $\Delta t_{out}$. This percentage is shown for different number of outbursts that occur within the 9 years of VVV observations. The analysis shows that, for both amplitude ranges, we are missing a large proportion of short-term YSOs outbursts, especially if there are fewer than 6 outbursts during VVV observations. For intermediate and long-term outbursts, the ability to detect them strongly depends on the amplitude of the outbursts. In both cases we are missing a few objects if the YSOs go through only one outburst and amplitudes are randomly distributed between 1.5 and 3 mag. If the outburst amplitudes are larger than these values, then we are only missing a small percentage of objects with $\Delta t_{out}$ between 1 and 2 years. For both low and high amplitudes we note a drop in the ability to detect long-term outbursts once $\Delta t_{out}$ approaches the duration of the survey.  This is not surprising given that if an outburst has $\Delta t_{out}$>9 yr and it happens before the start of VVV observations, we would not classify this YSO as a high-amplitude variable star. The modelling of Appendix \ref{app:a} takes into account this possibility and some synthetic outbursts can occur before the start of the VVV time-series.

The results from our analysis agree with the predictions from the simulated light curves. We find that all the objects that are classified as short-term show two or more outbursts. This implies that we are likely missing a large proportion of short-term outbursting YSOS that only have one outburst during VVV observations. The results of the simulated light curves shown in Fig. \ref{fig:ndet} show that we might be missing as much as 75\% of short-term outbursts and up to 30\% of intermediate type outbursts. Therefore we estimate that the number of YSOs detected in our analysis in Section \ref{sec:method} need to be corrected by factors of 4 and 1.4 for short-term and intermediate eruptive YSOs, respectively. In general, the long-term outbursts tend to have larger amplitudes than the other types of outbursts, which would imply that the outburst amplitudes in this class are intrinsically larger. If so, we are likely not missing a large sample of these objects and therefore we do not to correct for this effect.

The different definitions of the Class I stage can also have an effect on the estimated value of the incidence of episodic accretion. As mentioned in Section \ref{ssec:class} the colour-based classification of \citet{2009Gutermuth} and \citet{2014Koenig} includes objects with negative values of $\alpha$ (but larger than $-$0.3). The latter could be considered as Class II or Flat-spectrum (transition) YSOs. Objects with positive values of the spectral index are considered as embedded or Class I YSOs in the original definition from \citet{1987Lada}, but sources with $0 <\alpha<0.3$ can also  be defined as transition objects following the classification of \citet{1994Greene}. If we only consider YSOs with $\alpha>0$ as Class I YSOs, then our VIRAC2 detected sample reduces from 5661 to 3807 objects. Only including objects with $\alpha>0.3$ further reduces the sample to 2181 YSOs.

\subsection{Contamination}

Contamination by non-YSOs is expected in our catalogue of Class I YSOs. The contamination by extragalactic sources is not expected to be larger than 0.4\% in the red sources catalogue of \citet{2008Robitaille}, whilst \citet{2020Kuhn} also expect a low level of contamination of extragalactic sources due to the shallow IRAC observations in the GLIMPSE surveys. The mid-IR colours of known population of extragalactic surveys also follow the distribution of sources classified as likely contaminants in \citet{2020Kuhn}. Finally, \citet{2016Marton} indicates that the Class I/II candidates contain less than 1\% of contamination from extragalactic and Galactic sources. However, \citet{2020McBride} estimates that the contamination in the \citet{2016Marton} catalogue is much higher, with a large fraction of contamination from likely evolved stars.

We have stated in Section \ref{ssec:sample} that \citet{2008Robitaille} uses the 4.5 $\mu$m magnitude and [8.0]-[24] colour of intrinsically red objects to classify objects into either YSOs or evolved stars. To determine the contamination of our sample, we use a similar criteria to classify objects in our catalogue of 7205 Class I YSOs (Section \ref{ssec:class}). For sources arising from the \citet{2016Marton} catalogue we use the $W2$ and $W4$ filters\footnote{Inspection of {\it Spitzer} selected sources shows that objects with [8.0]-[24]$<$2.5 also show [4.5]-[24]$<$4.8 mag. Therefore we use the criterion W2-W4$<$4.8 to select likely contaminants}. We find that 635 objects out of 5661 VIRAC2 detected sources could be classified as evolved stars based on the criteria from \citet{2008Robitaille}, which amounts to a contamination of $\sim11.2\%$. 

We also need to consider the possibility that extinction can be driving the variability in some of our YSOs. As shown in Fig. \ref{fig:reddening} many of our eruptive YSOs have $\Delta$W2/K$_{\mathrm{s}}$ ratios that fall around or below the expectations from variability driven by changes in the extinction. If we assume that all objects below the \citet{2019Wang} line in Fig. \ref{fig:reddening} are being driven by extinction, then 20 out of 73 eruptive variable YSOs with valid WISE data, would be contaminating our sample. However, we notice that the $\Delta$W2/K$_{\mathrm{s}}$ ratios are uncertain due to WISE and VVV not being entirely contemporaneous surveys. In addition, being below the \citet{2019Wang} line does necessarily imply extinction-driven variability as four spectroscopically-confirmed eruptive YSOs fall below this line.

\subsection{Estimate of the incidence}

First we estimate the incidence of episodic accretion without taking into account the issues raised above. From the total sample of 5661 Class I YSOs with VIRAC2 counterparts, we find 304 objects that show high amplitude ($\Delta K>2$~mag) variability over the 9 years of VVV/VVVX observations. This represents 5.37\% of the Class I population. From these, 97 objects, or $\sim$1.71\% of Class I YSOs, are classified as eruptive variables. By dividing the sample of eruptive variables according to outburst duration, we find that 0.3\%, 0.65\% and 0.76\% of Class I YSOs show short-term, intermediate and long-term outbursts, respectively.

Table \ref{tab:incidence} shows the change in these values after taking into consideration the effects of the definition of the Class I stage from the spectral index, contamination of non-YSOs, contamination of other classes of variability, and the effect of our choice of amplitude cut. We find that depending of our choice of spectral index $\alpha$, the incidence of episodic accretion among Class I YSOs raises from 1.71$\%$ to between 2.36 and 2.78$\%$.

 \begin{table*}
	\centering
	\caption{Incidence of episodic accretion.}
	\label{tab:incidence}
\resizebox{\textwidth}{!}{
   	\begin{tabular}{lccccccc} 
		\hline
		Classification & N$_{Class~I}^{\dagger}$ & Outburst-type & N$_{outburst}$ & fraction$_{1}^{a}$($\%$) & fraction$_{2}^{b}$($\%$) & Final$^{c}$($\%$) & Total (($\%$))\\
		\hline
		$\alpha\geq-0.3$ &  5026 & Short-term & 17 & 0.34 & 0.25 & {\bf 0.98} & \\
              & & Intermediate & 37 & 0.74 & 0.53 & {\bf 0.75} & \\
             & & Long-term & 43 & 0.86 & 0.63 & {\bf 0.63} & \\
             & & & &  &  & & {\bf 2.36} \\
            \hline
            \hline
            $\alpha\geq0$ &  3380 & Short-term & 12 & 0.36 & 0.26 & {\bf 1.03} & \\
              & & Intermediate & 28 & 0.83 & 0.6 & {\bf 0.84} & \\
             & & Long-term & 24 & 0.71 & 0.52 & {\bf 0.52} & \\
             & & & &  &  & & {\bf 2.39} \\
             \hline
            \hline
            $\alpha\geq0.3$ &  1936 & Short-term & 10 & 0.52 & 0.38 & {\bf 1.5} & \\
              & & Intermediate & 15 & 0.77 & 0.56 & {\bf 0.79} & \\
             & & Long-term & 13 & 0.67 & 0.49 & {\bf 0.49} & \\
             & & & &  &  & & {\bf 2.78} \\

		\hline
		\multicolumn{7}{l}{$\dagger$ Corrected for possible contamination of non-YSOs.}\\
  \multicolumn{7}{l}{$a$ N$_{outburst}$/N$_{Class~I}$.}\\
  \multicolumn{7}{l}{$b$ fraction$_{1}$ corrected for contamination by extinction-driven variables.}\\
  \multicolumn{7}{l}{$c$ fraction$_{2}$ corrected for selection effects (amplitude cadence).}\\
	\end{tabular}}
\end{table*}
  
\section{Frequency of FUor outbursts}\label{sec:taucalc}

To estimate the frequency of outbursts, or recurrence timescales, \citet{2015Hillenbrand} and \citet{2019Contreras} assume low event rates for outbursts that do not repeat on observable timescales. This type of assumption only applies to objects classified as FUors. Hence we can provide an estimate on the recurrence timescale, $\tau$, for VVV objects classified as long-term eruptive variables.

\subsection{Derivation of $\tau$}

 Following \citet{2019Contreras} we derive the the probability density function $p(\tau|k)$ of a particular value of $\tau$, given the number of events $k$ as
 
\begin{equation}
p(\tau|k) ={{p(\tau)R(k|\tau)}\over{R(k)}}, 
\label{eq:eqn1}
\end{equation}

\noindent where $R(k|\tau)$ is given by a Poisson distribution

\begin{equation}
R(k|\tau) = {{(Nt/\tau)^k}\over{k!}}e^{-Nt/\tau},
\label{eqn:poisson}
\end{equation}

with $N$ the total number of objects in our sample and $t$ the time baseline of observations. By evaluating $R(k)$ assuming a prior with $p(\tau)\propto 1/\tau$, i.e a flat distribution in $ln(\tau)$, yields 
\begin{equation}
\label{eqn:or8}
p(\tau | k) ={{(Nt)^k}\over{(k-1)!\,\tau^{k+1}}}{e^{-Nt/\tau}},
\end{equation}

Differentiating this to find the turning point, yields the most probable value of $\tau$ as
\begin{equation}
\tau={{Nt}\over (k+1)}.
\end{equation}

The confidence limits of $\tau$  are given by the values of $\tau$ which enclose the most likely 68 percent of the probability given by Equation \ref{eqn:or8}.
We integrated Equation \ref{eqn:or8} over all values of $\tau$ where $p$ exceeded a given value $l$. The value $l$ is then decreased until the integral reaches 0.68, at this point  the extreme values of $\tau$ represent the 68 precent confidence limits.

\begin{figure}
	\resizebox{\columnwidth}{!}{\includegraphics[angle=0]{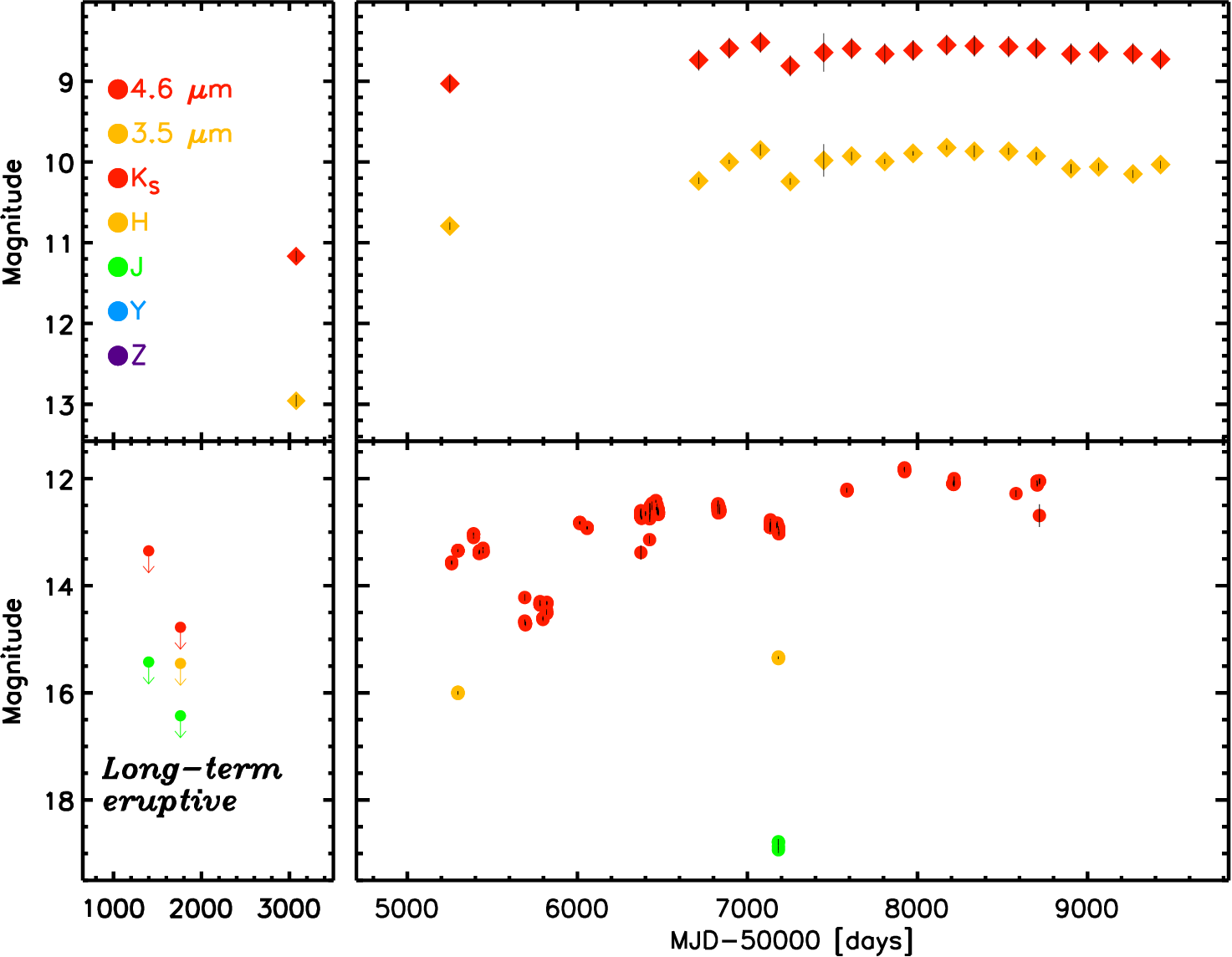}}
	 \caption{Example of YSO1956 , where the inspection of long-term data from e.g. 2MASS, DENIS and {\it Spitzer} allows us to classify this object as a long-term outburst. In the figure smaller solid circle with downward arrows represent 5$\sigma$ limits from DENIS and 2MASS surveys.}
    \label{fig:ex_fader}
\end{figure}

We find 43 objects that are classified as long-term outbursts over a total sample of $N=5661$~YSOs. To estimate $\tau$ we need to define a time baseline, $t$. The selection of objects is primarily based in the fact that they display $\Delta K>2$~mag over the VVV observations, which cover 9 years. However, comparison with 2MASS, DENIS and/or {\it Spitzer} observations was also used to classify some YSOs as long-term eruptive variables. For example, YSO1956 (Fig \ref{fig:ex_fader}) is selected as a high-amplitude variable YSO, however, based on VVV observations alone, this object resembles more a dipper than an eruptive YSO. Increasing the baseline to cover {\it Spitzer}, DENIS and 2MASS observations shows that the object went into outburst somewhere between 2004 and 2010. Since the comparison with past near- and mid-IR surveys is done after the selection of high-amplitude variables from VVV, we are likely to miss detecting objects that went into outburst before the start of VVV observations in 2010 and that did not display large variability over the timescales covered by the survey. To study the effects in the choice of $t$, we determine $\tau$ using $t=9$, $t=19$ and $t=14$ years. The latter value is simply taken the mean between the 9 years covered by VVV and 19 years once we include 2MASS/DENIS.

\begin{figure*}
	\resizebox{1.5\columnwidth}{!}{\includegraphics[angle=0]{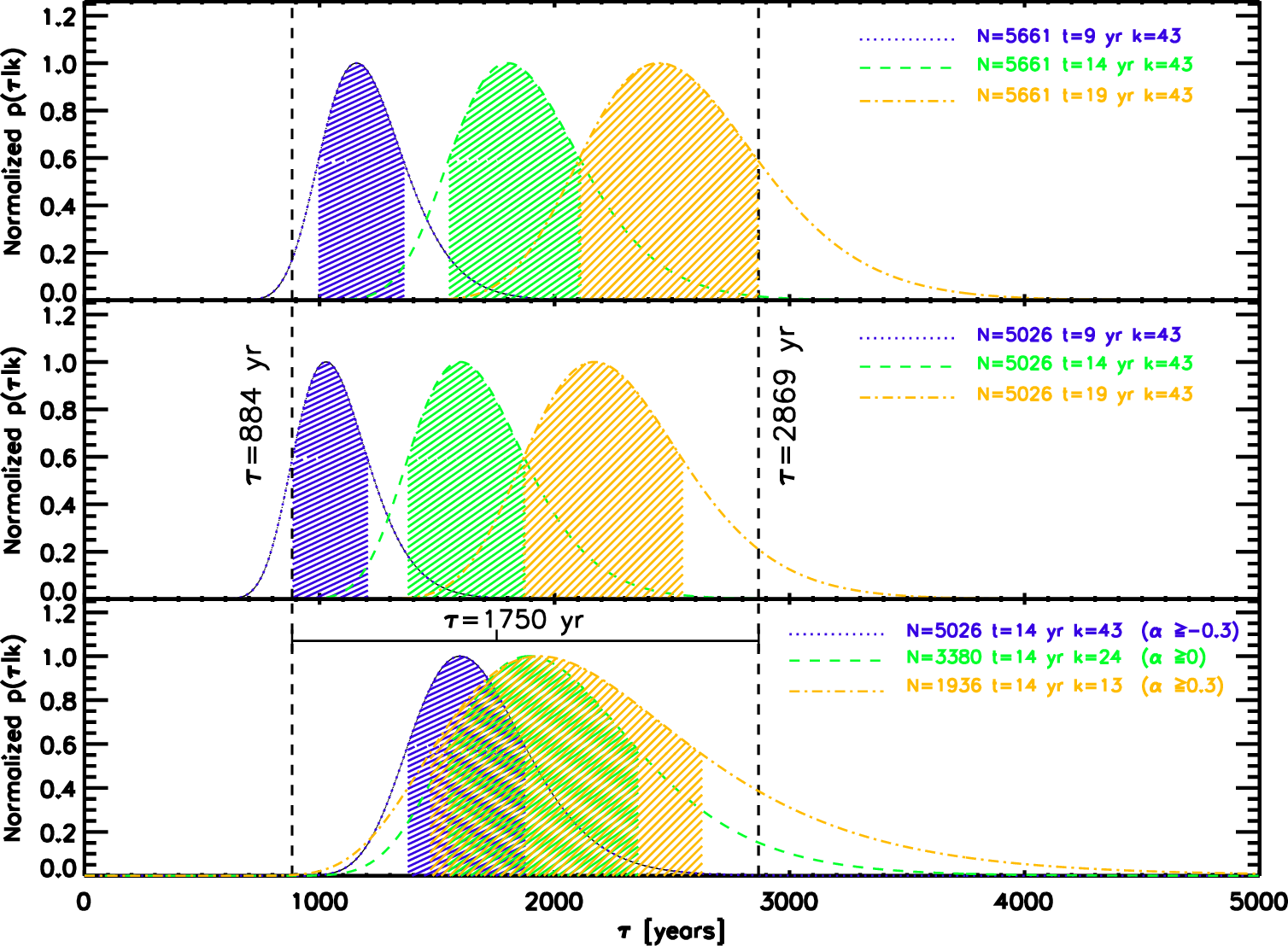}}
	 \caption{Estimates of $\tau$ for different values of the parent sample $N$, time baseline, $t$ and number of outbursts, $k$, as explained in the main text. Confidence intervals are shown as blue, green and orange dashed lines. Depending on the values of $N, t$, and $k$ the confidence intervals are always contained between $\tau=884$ and $\tau=2869$ years.}
    \label{fig:tau}
\end{figure*}

In Fig. \ref{fig:tau} we show the results of estimating $\tau$ when assuming different values of $t$. The most probable value of $\tau$ does change depending on the choice of $t$, however it consistently moves between $\approx$ 1000 and 3000 yr. A similar effect is observed by considering the contamination from evolved stars estimated in Section \ref{sec:inc_out}, which reduces the total sample of YSOs to $N=5026$ objects.

We finally consider the different definitions of the Class I stage (see Section \ref{sec:inc_out}). Assuming a value of the time baseline of $t=14$~yr, and using $N=5026$ after considering contamination, we investigate how $\tau$ changes for different definitions of the Class I stage, i.e YSOs with $\alpha \geq -0.3$ (or full sample), $\alpha \geq 0$ \citep[original][definition]{1987Lada} and $\alpha \geq 0.3$. The bottom plot of Fig. \ref{fig:tau} shows that the value of $\tau$ for the different samples are similar within the confidence intervals.

Finally, we estimate the recurrence timescales of FUor outbursts in the Class I stage as the average of the most probable values of $\tau$ of the different selections of $\alpha$, $N$, $k$ and $t$, yielding a value of $\tau=1.75$~kyr. The lower and upper limits are taken as the 884 and 2869 yrs, respectively (shown in Fig. \ref{fig:tau}). This yields a final value of $\tau=1.75^{+1.12}_{-0.87}$~kyr.

\subsection{Previous estimates of $\tau$}

\citet{2013Scholz}  searched for young eruptive variables by comparing {\it Spitzer} with {\it WISE} photometry, thus providing a 5 year baseline. The authors determine that YSO outbursts occur between every 5000 and 50000 yr. However, the YSO sample in \citet{2013Scholz} is not divided into different evolutionary classes. By re-classifying the YSO samples A and B of \citeauthor{2013Scholz}, \citet{2019Contreras} determines frequencies of FUor outbursts in the Class I stage of $3.0^{+12.7}_{-1.9}$~kyr  and $6.0^{+25.3}_{-3.9}$kyr for samples A and B, respectively.

In the search for Class II Outbursts, \citet{2019Contreras} find that the overall sample of Class II YSOs, defined through colour-colour diagrams and spectral indices, suffers from contamination from Class I YSOs. \citet{2019Contreras} finds three YSO outbursts that are more likely to be Class I YSOs over a total sample of 1043 Class I contaminants. Using these values, \citet{2019Contreras} find an outburst interval of $13^{+15}_{-6}$\,kyr (68 percent confidence) in Class I YSOs. 

\citet{2019Fischer} studied the large amplitude mid-infrared variability for a sample of 319 protostars (YSOs with class designations 0, I or flat) in Orion, by comparing {\it Spitzer} versus {\it WISE} photometry. Based on the recovery of two outbursts over the 6.5 yr baseline \citet{2019Fischer} determine an outburst rate of 1000 yr with a 90\% confidence interval of 690 to 40300 yr. A similar analysis of the variability protostars  (YSOs with class designations 0, I or flat-spectrum) in nearby star forming regions using 6.5 yr of NEOWISE observations by \citet{2021Park}, finds an outburst rate of 683 yr with a 90\% confidence interval of 356 to 1579 yr, in line with the results from \citet{2019Fischer}.  Finally, \citet{2022Zakri} analysed the light curves of Class 0 YSOs in the Orion clouds using {\it Spitzer}/IRAC photometry spanning from 2004 to 2017. The detection of three outburst leads to a  recurrence timescale of 438 yr with a 95\% confidence interval of 161 to 1884 yr. 

\citet{2019Hsieh} observes a sample of 39 Class 0 and I YSOs in the Perseus molecular cloud. From ALMA observations, \citet{2019Hsieh} are able to trace the location of CO and H$_{2}$O snowlines in the YSO sample. Accretion outbursts can move the location of the snowline towards larger radii \citep{2016Cieza}. Therefore, if the snowline is determined to be located at larger radii than expected from the current luminosity of the YSO, this implies a recent outburst. \citet{2019Hsieh} determine that the interval between outbursts is $\sim$2400 yr and $\sim$8000 yr for the Class 0 and Class I stages, respectively. In a similar, previous analysis of SMA observations of the C$^{18}$O emission in 16 protostars, \citet{2015Jorgensen} yields a rate of about 1 outburst every 20000 yr.  

The observation of emission knots in jets from YSO outflows are likely associated with past accretion events \citep{2012Ioannidis}. The observation of gaps between H$_{2}$ knots indicate a time between ejection events of $\simeq1000$~years \citep{2012Ioannidis, 2016Froebrich, 2018Makin}. The H$_{2}$ jets likely trace the accretion related events during the earlier stages of young stellar evolution \citep{2012Ioannidis}

\begin{figure*}
	\resizebox{1.5\columnwidth}{!}{\includegraphics{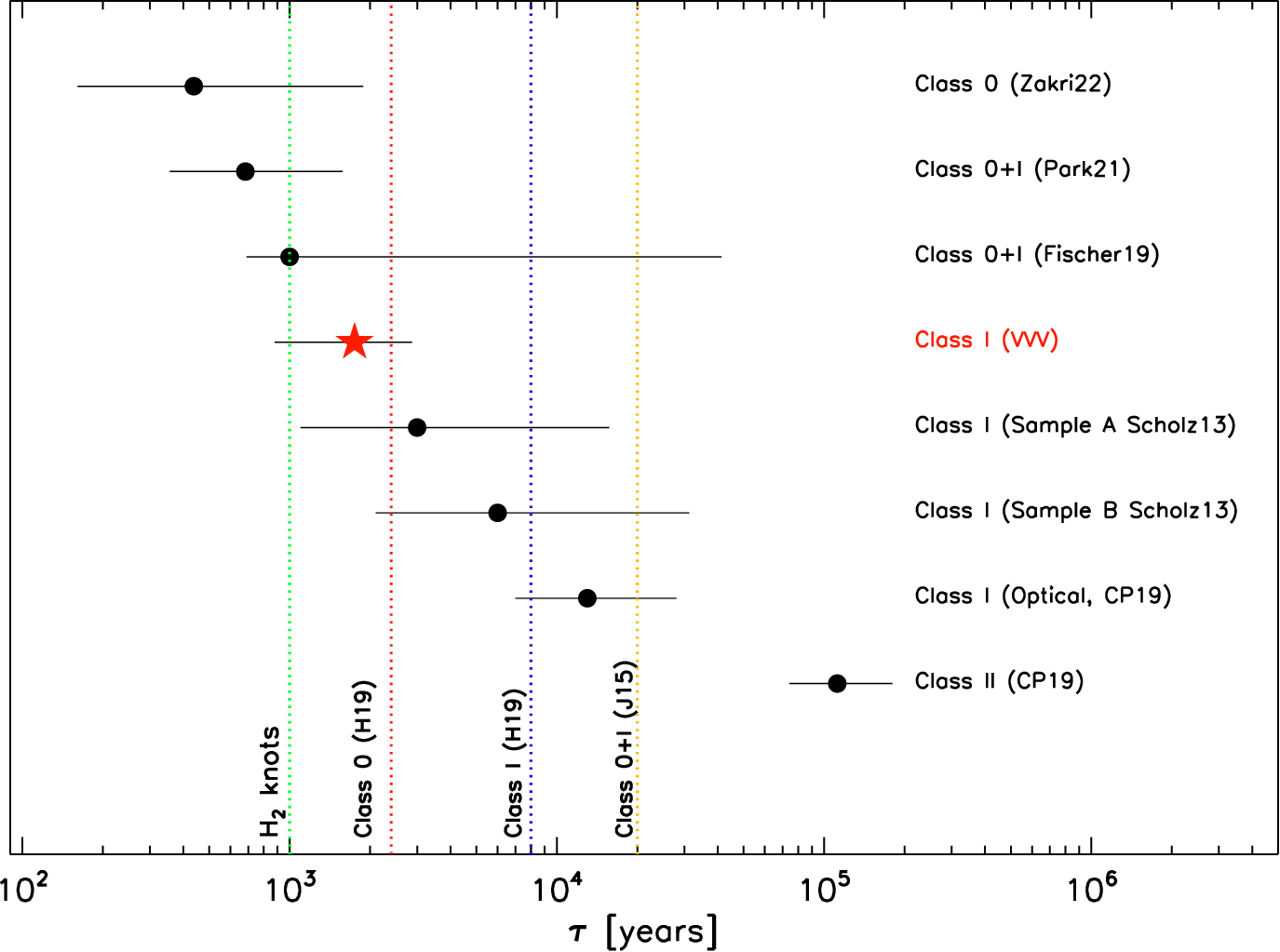}}
	 \caption{Comparison of the results from this work with the different estimates of $\tau$ from the literature, as described in the main text.}
    \label{fig:tauc}
\end{figure*}

\subsection{Discussion on $\tau$}

One major caveat of any discussion on the frequency of FUor outbursts is the assumption that all YSOs goes through these episodes of high accretion.  Millimetre continuum observations with ALMA have shown that FUor disks are more massive and compact than the disks of other classes of eruptive variables \citep{2018Cieza}, and those of regular Class II and Class I YSOs \citep{2021Kospal}. This leads to the possibility that not all YSOs gain their mass through episodic accretion, but FUors are instead a different type of YSOs that follow a particular path in their evolution that leads to episodes of high accretion \citep{2022Fischer}.

\begin{figure}
       \resizebox{\columnwidth}{!}{\includegraphics{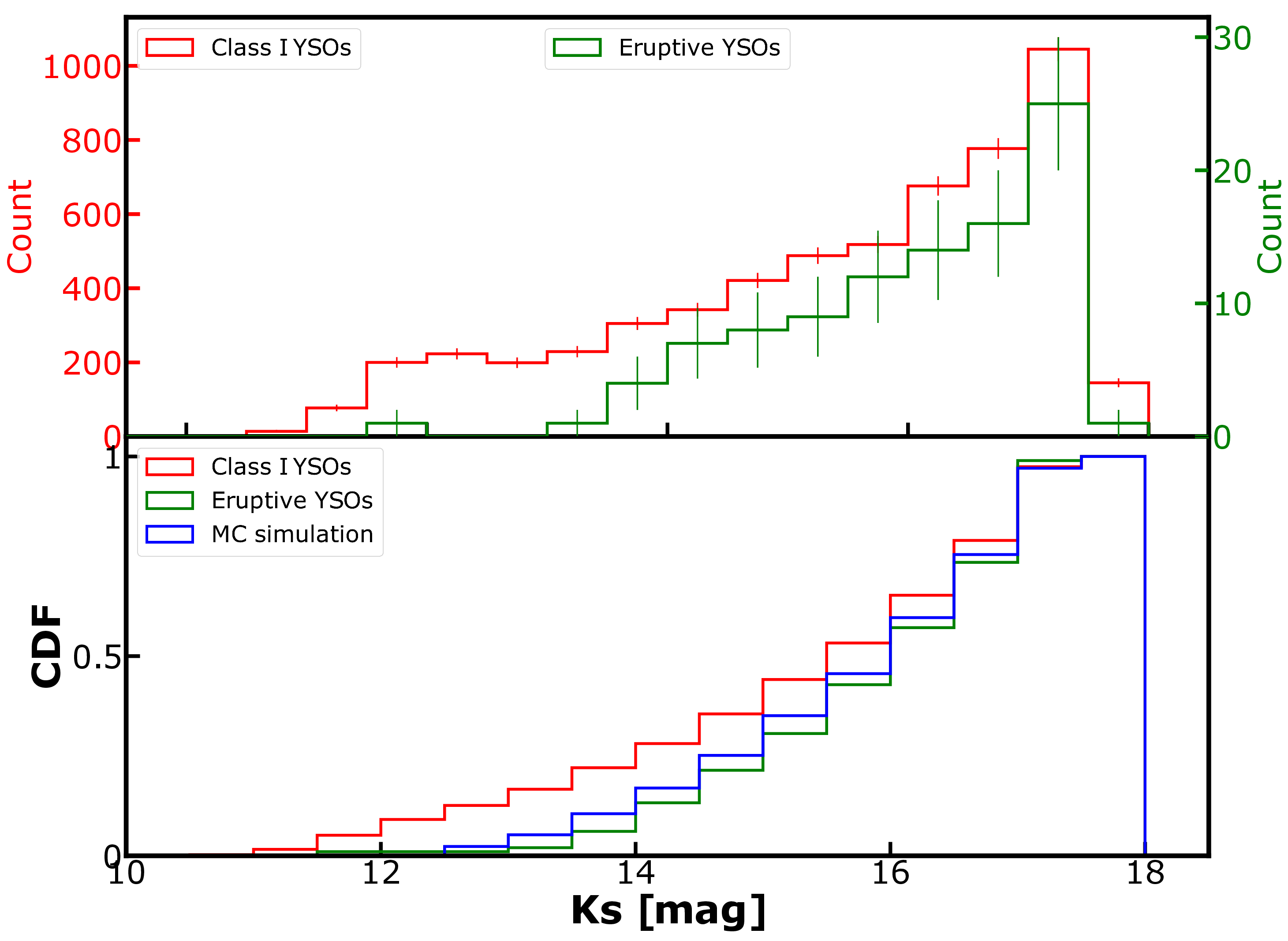}}
         
	 \caption{(Top) $K_{\rm s}$ magnitude  distribution of the faintest epoch in the light curve for the parent sample of Class I YSOs (solid red line), and for accretion driven outbursts (solid green line). (Bottom) Cumulative distribution function of the faintest $K_{\rm s}$ magnitude in each light curve, shown for Class I YSOs (red), accretion driven outbursts (green) and simulated outbursts in the Class I YSOs (blue), accounting for saturation (see main text).}
    \label{fig:mbias}
\end{figure}

If it is assumed that all YSOs go through these episode of high accretion, there are other caveats that need to be discussed. Every estimate of $\tau$ arising from the detection of FUor outbursts from a sample of YSOs, and over a defined time baseline suffers from similar uncertainties \citep{2022Fischer}. Firstly, defining the parent sample, $N$, depends on the classification criteria, and contamination from evolved stars can be significant. In addition, there are several issues that can lead to uncertainties in the number of outbursts detected ($k$) in the calculations of Section \ref{sec:taucalc}. There are difficulties in distinguishing true outbursts from other mechanisms that drive long-term, large amplitude variability in YSOs (such as extinction) or in contaminating objects such as AGBs. We could also be missing eruptive YSOs that are slowly fading back into quiescent stages, and therefore do not vary by more than our amplitude cut of 2 magnitudes over the VVV coverage (we are investigating these type of objects in a separate publication, Contreras Pe\~{n}a et al, in prep.). Only six {\it Faders} from our high-amplitude sample (Section \ref{ssec:hamp_class}) are not classified as accretion driven outbursts (see Section \ref{sec:acc_out}) as these do not show high-amplitude variability in the mid-IR.   

The increase of luminosity with distance observed in Fig. \ref{fig:luminosity} could indicate that the sample of accretion-driven outbursts is affected by Malmquist bias. If outbursts are systematically brighter than the non-eruptive population then we could detect accretion-driven outburst at larger distances and therefore overestimate the frequency of these events. Unfortunately, {\it Gaia}-based distances are only available for a small number of sources in the Class I YSO sample, and even then, these estimates are strongly biased towards nearby sources with low extinction. To investigate the likelihood of Malmquist bias affecting our results, we compare the distribution 
 of the $K_{\rm s}$ magnitude of the faintest epoch for each source in the parent sample of Class I YSOs with the same distribution for the accretion-driven outbursts (top panel of Fig. \ref{fig:mbias}). Figure \ref{fig:mbias} shows no indication that accretion-driven outbursts are brighter than the parent sample. A similar conclusion is reached when comparing the cumulative distribution function (CDF) of the faintest K$_{\rm s}$ magnitude in the light curves for all sources with the same distribution for accretion driven outbursts (see bottom panel of Fig. \ref{fig:mbias}). These figures indicate that the accretion-driven outbursts share similar brightness distribution with their parent Class I sample towards the fainter end. However, there is an obvious difference between the distributions when K$_{\rm s}$ is brighter than 13.5 mag. This differences is probably due to the fact that sources with quiescent magnitudes brighter than  K$_{\rm s}\sim$13~mag will be above the saturation limit of the VIRAC2 light curves (K$_{\rm s}\sim$~11~mag) during outburst and therefore are less likely to be detected by \textsc{DoPhot}. There may also be a small effect of contamination by  evolved giant stars \citep[as observed in][]{2017Contreras_a,2017Contreras,2021Guo}.


To test the effect of the VIRAC2 bright limit on the observed distribution of eruptive YSOs, we performed a simple analysis. First we take 200 YSOs that follow the same distribution in K$_{\rm s}$ as the parent Class I sample. Then for an individual object in each K$_{\rm s}$ bin we:

\begin{itemize}
    \item[(a)] Assume a random quiescent magnitude, K$_{\rm s,q}$, but within the magnitude range of the bin from which the source was taken.
    \item[(b)] Assume that this YSO goes into outburst with a random amplitude, $\Delta$, between 2 and 4 mags ($\sim90\%$ of the eruptive YSO sample have these values in amplitude). Then the bright magnitude, K$_{\rm s,b}$, is taken as K$_{\rm s,b}$=K$_{\rm s,q}-\Delta$.
     \item[(c.1)] If $11>$K$_{\rm s,b}>10$ then in 15\% of the cases we assume that the object is not detected and we set K$_{\rm s,b}=$K$_{\rm s,q}$. In the other 85\% of the cases, we assume detection and the value of K$_{\rm s,b}$ is not corrected.
     \item[(c.2)] If $10>$K$_{\rm s,b}>9$ then in 57\% of the cases we assume that the object is not detected and we set K$_{\rm s,b}=$K$_{\rm s,q}$. In the other 43\% of the cases, we assume detection and the value of K$_{\rm s,b}$ is set at K$_{\rm s,b}=10.5$.
     \item[(c.3)] If K$_{\rm s,b}<9$ then in 79\% of the cases we assume that the object is not detected and we set K$_{\rm s,b}=$K$_{\rm s,q}$. In the other 21\% of the cases, we assume detection and the value of K$_{\rm s,b}$ is set at K$_{\rm s,b}=10.5$.\footnote{ The percentages in items (c.1) through (c.3) are estimated by comparing the number of objects in the Class I sample with 2MASS K$_{\rm s}$ magnitudes, that have or lack VIRAC2 detections. YSOs with 2MASS K$_{\rm s}<10$ mag saturate in VIRAC2, and when detected tend to show values of K$_{\rm s}\sim10.5$ in VIRAC2 photometry. Thus we choose this limit for items (c.2) and (c.3).}
     \item[(d)] The new amplitude is taken as K$_{\rm s,q}-$K$_{\rm s,b (corrected)}$. We mark the object as detected as an outburst if this new amplitude is larger than 2 mags.
    \item[(e)] Estimate a new distribution of eruptive YSOs from the detected objects.
     \item[(f)] Repeat this 1000 times, and the final distribution is assumed as the mean of the counts in each bin.
\end{itemize}

The simulated distribution of eruptive YSOs is shown in the bottom panel of Fig. \ref{fig:mbias}. The latter looks very similar to that of the observed sample of eruptive YSOs, demonstrating that the saturation limit of VIRAC2 plays an important role in the observed distribution.

In spite of these caveats, the comparison of different measures of $\tau$ across the different classes (or stages) of young stellar evolution (Fig. \ref{fig:tauc}) seems to point to an increase in the frequency of these events for younger stages of evolution. 

This is in agreement with the expectations from theoretical models. For example, \citet{2014Bae} find that gravitational instability(GI)-induced spiral density waves can heat the inner disc and trigger magneto-rotational instabilities (MRI) that will lead to accretion outbursts (or GI+MRI mechanism). They find that this process is more efficient at earlier stages when the envelope dominates the disc, with the number of outbursts decreasing as the star approaches the post-infall phase. Based on figure 3 of \citet{2014Bae}, \citet{2015Hillenbrand} estimate an outburst rate of $8\times10^{-5}$~year$^{-1}$~star$^{-1}$ during the class I stage, and $3\times10^{-6}$~year$^{-1}$~star$^{-1}$ during the class II stage, or a recurrence timescale of 12.5 and 333 kyr respectively. Finally, the hydrodynamical simulations of \citet{2021Riaz} show that embedded (or Class 0) YSOs emerging from protobinary configurations display strong accretion outbursts with recurrence timescales of 1 kyr.

\section{Summary}\label{sec:sum}

We have constructed a sample of 7205 Class I YSOs from {\it Spitzer} and {\it WISE} mid-IR photometry. Crossmatch with the VVV/VIRAC2 allowed us to investigate the near-IR variability of 5661 YSOs in our sample. The search for high-amplitude ($\Delta$K$_{s} >$2~mag) variability yields 304 sources, which showed a varied behaviour over the nine years of coverage by the VVV survey. Collection of additional near- to mid-IR information, as well as visual inspection of images and light curves, allowed us to classify 97 objects as candidate eruptive variable YSOs. From these, 70 objects are new discoveries, while 27 objects have been previously confirmed as eruptive variables through photometric and  spectroscopic observations.

The long-baseline and cadence of VVV observations allowed us to divide the candidate eruptive YSOs into three different classes according the duration of the outbursts. The YSOs are divided as short-term, intermediate or long-term, if the outburst duration is shorter than 1 year, between 1 and 9 years, and longer than 9 years, respectively. We find 43, 37 and 17 long-term, intermediate, and short-term outbursts, respectively.

We determine distances and luminosity for the sources classified as eruptive in our sample. We find that the luminosity reached during outburst is similar for all of the classes (short-term, intermediate and long-term).

From the sample of eruptive YSOs we are able to estimate the incidence of episodic accretion during the Class I stage. This number varies between 1.7 and 2.8$\%$ as we also consider the effects of  e.g. contamination from non-YSOs or the definition of the Class I stage from the spectral index $\alpha$.

\citet{2009Enoch} finds the majority of Class I YSOs in nearby star forming regions have luminosities of $L<1 L_{\odot}$. Only 5$\%$ of their sample are found at the higher-end of the luminosity distribution corresponding to stars accreting at $\dot{M}\sim 10^{-5}$~M$_{\odot}$~yr$^{-1}$. The latter value of $\dot{M}$ is consistent with the expected value of the accretion rate reached during an FUor (or long-term) outburst. This could be an argument to support that the 5$\%$ estimate of \citet{2009Enoch} applies only to FUor outbursts. However, our results show that eruptive variables can reach similar luminosities, regardless of the duration of the outburst. Hence, our estimated value of the total incidence of episodic accretion of 2.8$\%$ is consistent with the observations of \citet{2009Enoch}.

The effect of episodic accretion on chemistry and the location of the snowline of various ices has been always been discussed in the context of FUor (long-term) outbursts as these are thought to be more extreme and reach higher luminoisities during outburst. The fact that we find that YSOs can reach high luminosities regardless of the length of the outburst could have interesting implications for planetary formation. EX Lupi-type events (i.e. short-term outburst) appear to contribute to the build-up of the crystaline dust component ubiquitously seen in comets \citep{2019Abraham,2023Kospal}. High luminosity outbursts move the location of the snowline towards large radii \citep{2016Cieza}, therefore repetitive, short-term (or intermediate), high-luminosity outbursts are bound to have an effect on the final composition of planets.

Finally, from 43 objects that are classified as long-term, we are able to determine the recurrence timescale, $\tau$ of FUor outbursts. Taking into account non-YSO contamination and the time baseline of our observations, we determine a value of $\tau=1.75^{+1.12}_{-0.87}$~kyr. Comparison with other estimates of $\tau$ for Class O, Class I and Class II YSOs from the literature, agrees with the expectation of a higher frequency of these type of outbursts towards younger evolutionary stages.

\section*{Acknowledgements}

This research has made use of the NASA/IPAC Infrared Science Archive, which is funded by the National Aeronautics and Space Administration and operated by the California Institute of Technology.

CCP was supported by the National Research Foundation of Korea (NRF) grant funded by the Korean government (MEST) (No. 2019R1A6A1A10073437)
ZG is supported by the ANID FONDECYT Postdoctoral program No. 3220029, and acknowledges support by ANID, -- Millennium Science Initiative Program -- NCN19\_171.
CCP and PWL also received some funding from grant ST/R000905/1 of the UK 
Science and Technology Facilities Council.



\section*{Data Availability}

The data underlying this article are available in the article and in its online supplementary material.

The $K_{s}$ light curves of various sources will be shared on reasonable request to the corresponding author.



\bibliographystyle{mnras}
\bibliography{ref_class1.bib} 




\appendix

\section{Synthetic light curves}\label{app:a}

\begin{figure}
	\resizebox{\columnwidth}{!}{\includegraphics{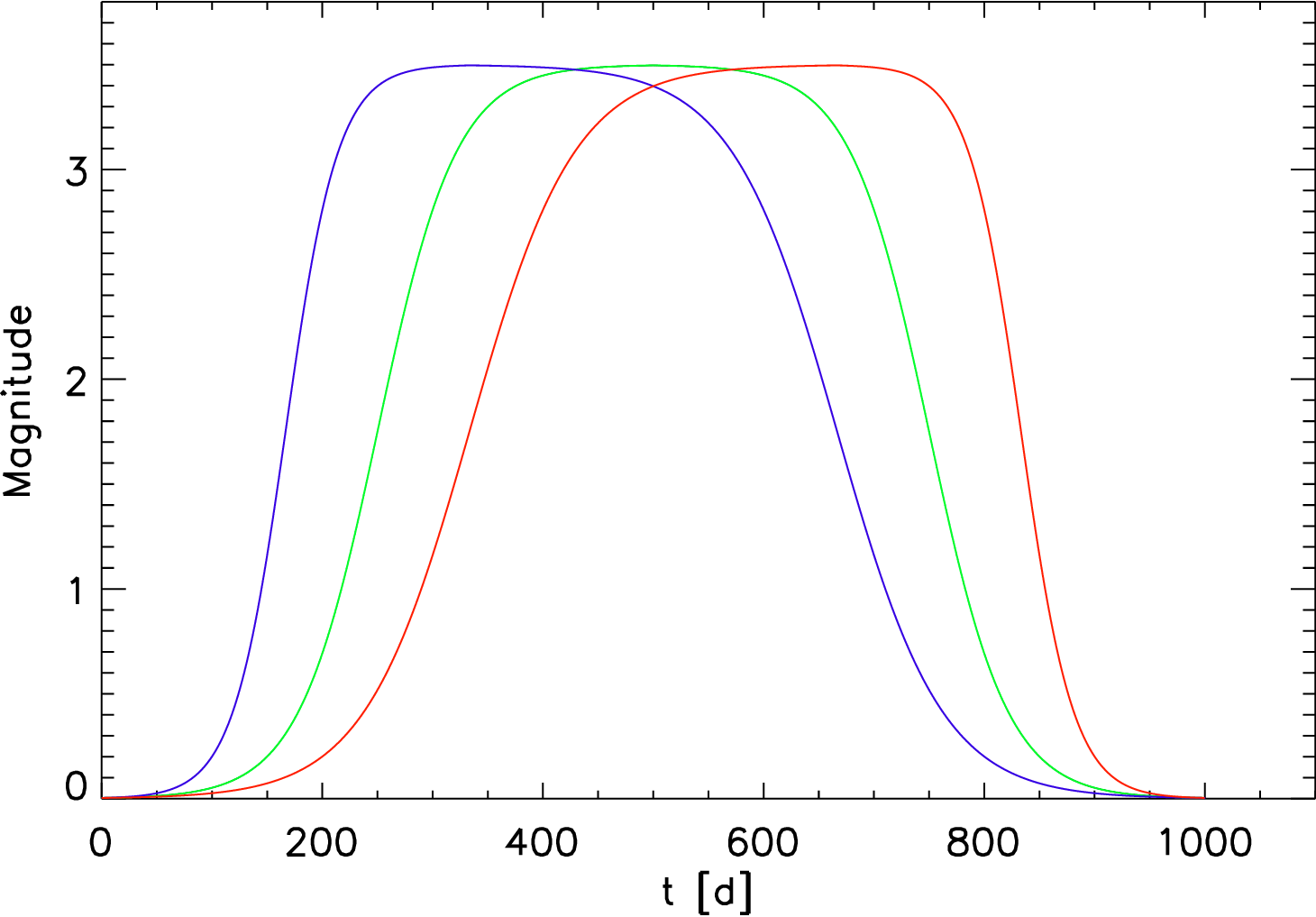}}
	 \caption{Type of synthetic outbursts created for this analysis.}
    \label{fig:app_f1}
\end{figure}

We want to determine whether the sampling of the VVV survey affects our ability to detect eruptive YSOs. To check for this, we generate synthetic light curves of eruptive YSOs. The outbursts are generated using two sigmoid functions. These model the rise and decline over times $t_{r}$ and $t_{d}$, respectively, of an outburst with duration $\Delta t_{out}=t_{r}+t_{d}$. The sigmoid functions are defined as

\begin{eqnarray}
S(t)&=&\frac{A}{1+e^{-k_{r}(t-t_{r,0})}},     0<t<t_{r}\\
S(t)&=&\frac{A}{1+e^{k_{d}(t-t_{d,0})}},     t_{r}<t<\Delta t_{out},
\label{eq:lc1}
\end{eqnarray}

with A the amplitude of the outburst, $t_{r,0}=t_{r}/2$, $t_{d,0}=t_{r}+t_{d}/2$, $k_{r}=7/(t_{r}/2)$ and $k_{d}=7/(t_{d}/2)$. The value of 7 in the exponents $k_{r,d}$ is chosen to ensure that the sigmoid functions reach $S(t=t_{r})=A$ or $S(t=\Delta t_{out})=0$ within the time spend in rise or decline, respectively. The outbursts are modelled as being symmetrical with $t_{r}=t_{d}=(\Delta t)/2$ or asymmetrical with either faster rise, $t_{r}=(\Delta t)/3$ and $t_{d}=2(\Delta t)/3$, or fast decline, $t_{r}=2(\Delta t)/3$ and $t_{d}=(\Delta t)/3$. The outburst type (symmetrical or asymmetrical) is chosen randomly before generating the outburst. Fig. \ref{fig:app_f1} shows an example of the types of symmetrical and asymmetrical outbursts generated in our models.

The first outburst in the synthetic light curve can occur at any time in the first 365 days. Once the first outburst is generated, the next outburst can occur within a time interval given by $\Delta t_{int}$. The probability, $p$, of generating an outburst within $\Delta t_{int}$ is given by a random distribution multiplied by a sigmoid function that approaches unity as $\Delta t_{int}$ is approached. If at any point within $\Delta t_{int}$, the probability $p$ is larger than 0.9 the next outburst is generated. This step is repeated until the synthetic light curve reaches 4500 days. 

To reproduce the cadence of the VVV survey, we randomly select the light curve of one of the 174 eruptive YSOs detected in our work. From this file we select the modified Julian day of $K_{\rm s}$ observations and subtract 55200 days. We generate a synthetic observed light curve by selecting the magnitudes from closest data points in the synthetic light curve to  the MJD-55200 array. If the maximum minus minimum magnitude in the observed light curve is larger or equal than 2 magnitudes, we mark it as a detection. The step of generating a synthetic light curve and obtaining observations from a random VVV sampling is repeated 200 times. 

To test low amplitudes range of outbursts, synthetic light curves are generated with amplitudes randomly selected within the range 1.5 to 3 mag. The second set of light curves are generated to study high-amplitude outbursts by randomly selecting amplitudes in the range between 3 to 5 mag.

In the analysis of low and high-amplitude ranges we used values for $\Delta t_{out}$ of 30, 90, 180, 360, 720, 1400, 2100, 2800 and 3500 d and for intervals, $\Delta t_{int}$ of 90, 180, 360, 720, 1400, 2100, 2800, 3500, 4200, 5000, 6000, 7000, 8000, 9000, 10000, 110000, 120000 d. For a given $\Delta t_{out}$, light curves are generated always using values of  $\Delta t_{int}$ that are larger than $\Delta t_{out}$.

\begin{figure*}
	\resizebox{\columnwidth}{!}{\includegraphics{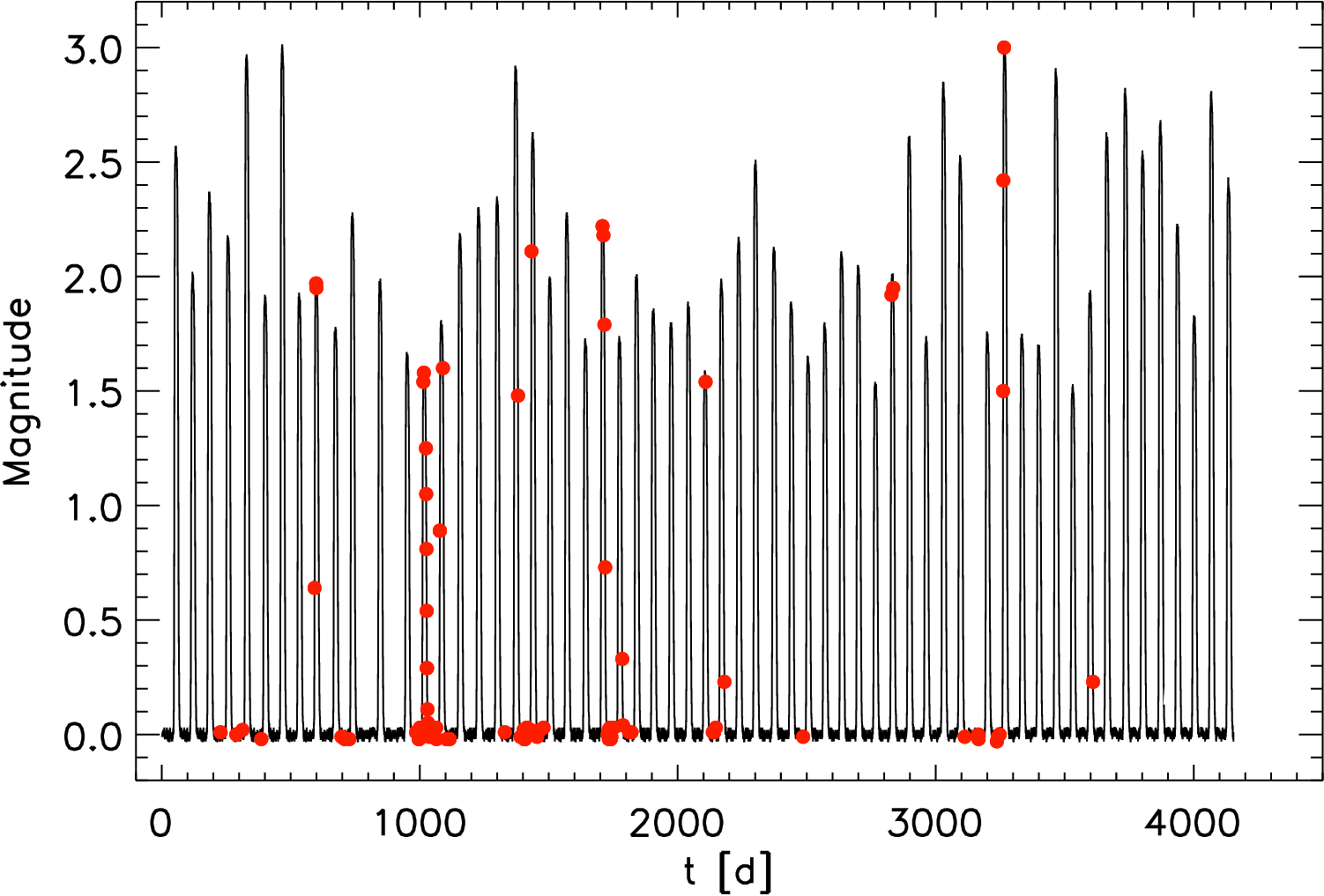}}
	\resizebox{\columnwidth}{!}{\includegraphics{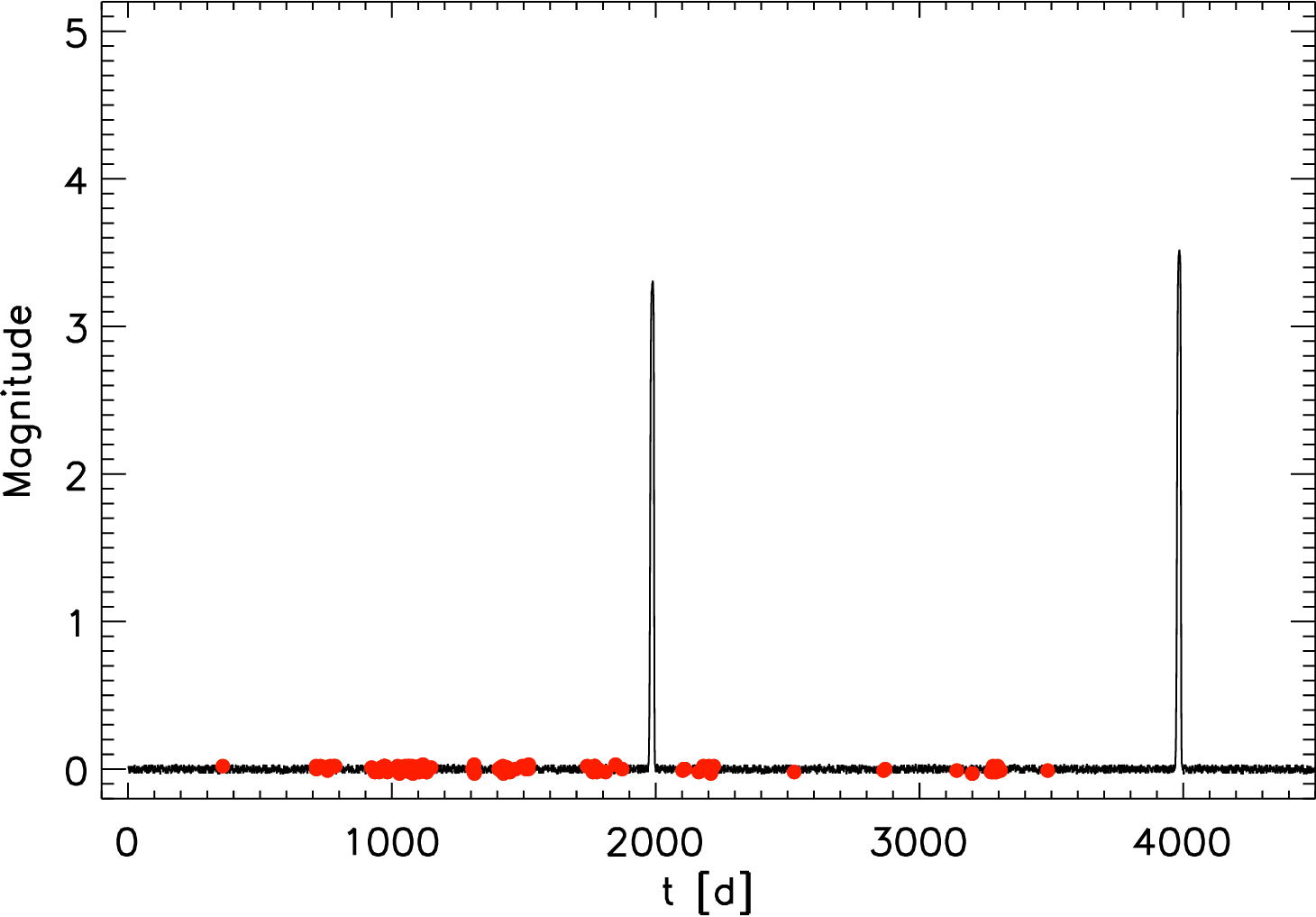}}\\
       \resizebox{\columnwidth}{!}{\includegraphics{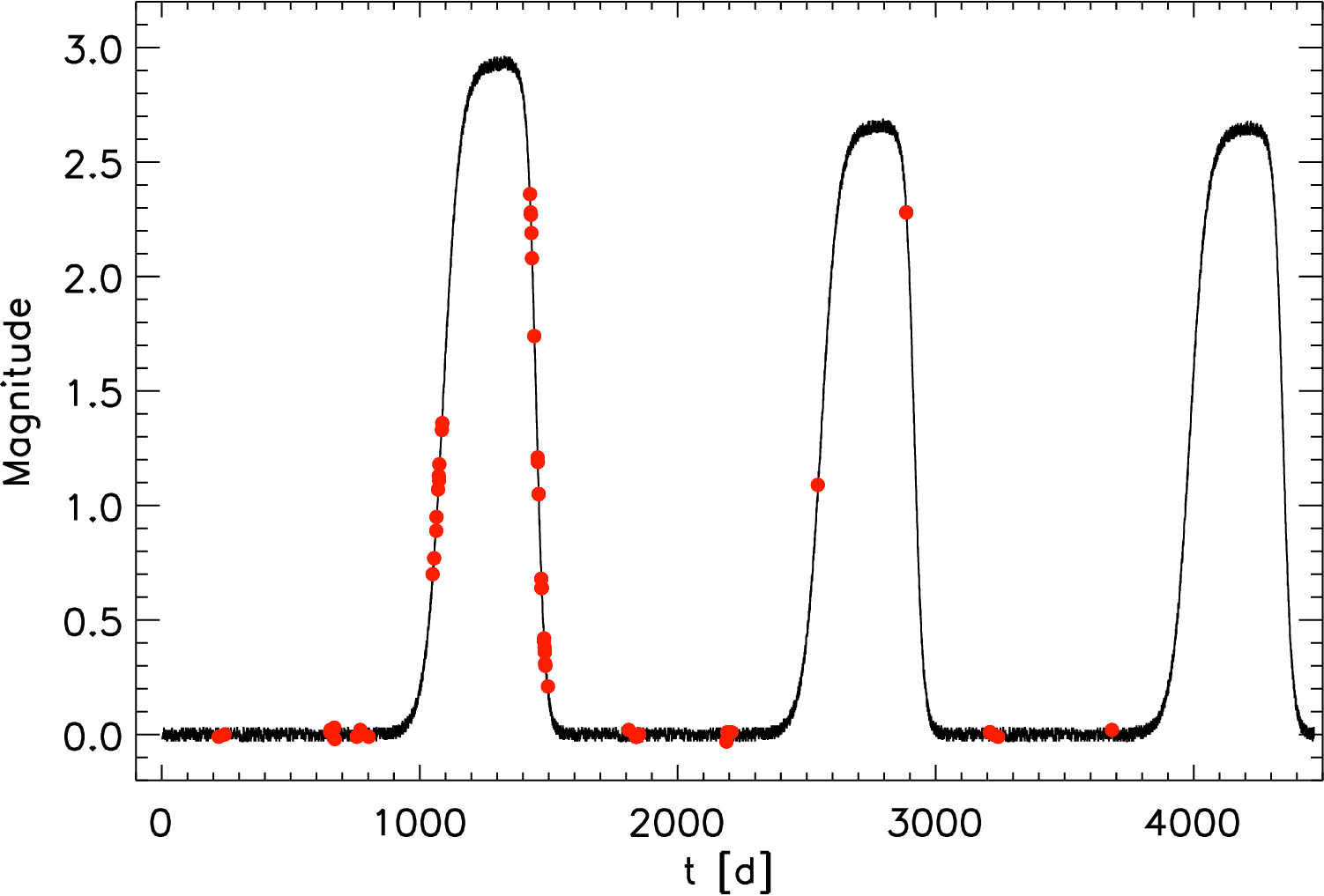}}
       \resizebox{\columnwidth}{!}{\includegraphics{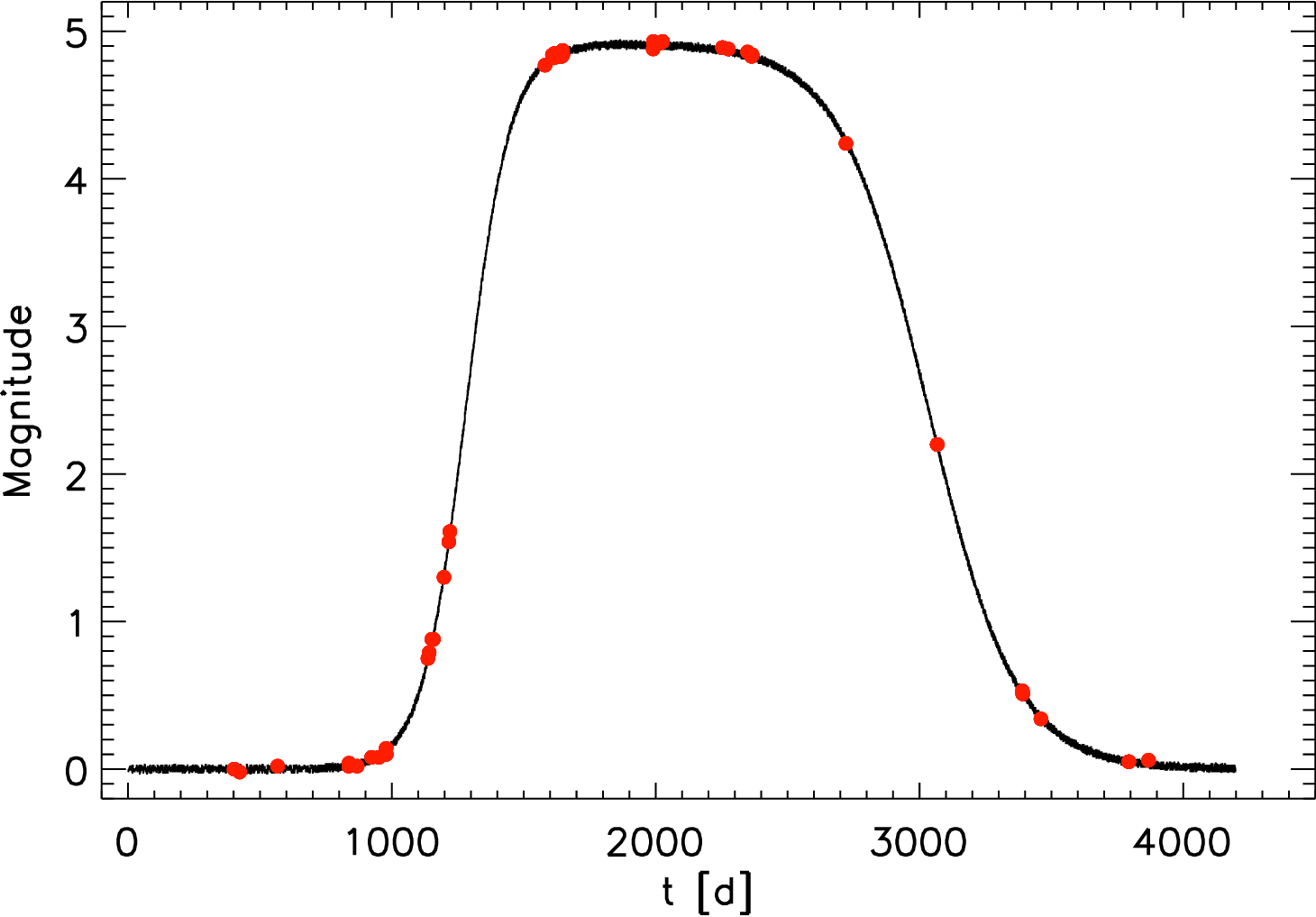}}
	 \caption{Examples of different synthetic light curves. We show short-term with numerous outbursts (top-left), short-term with only two outbursts in nine years (top-right), intermediate with three outbursts in nine years (bottom-left) and one long-term outburst (bottom-right). In the figures, red circles mark the epochs where these light curves would be observed by the VVV survey.}
    \label{fig:app_f2}
\end{figure*}

Fig. \ref{fig:app_f2} shows a few examples of synthetic and observed light curves for different values of the outburst amplitude, $\Delta t_{out}$ and $\Delta t_{int}$.

\bsp	
\label{lastpage}
\end{document}